\documentclass[]{article}

\usepackage{amssymb,bm,amsmath}
\usepackage{color, graphicx}
\usepackage{soul, url}
\usepackage{lineno}
\usepackage{authblk}

\usepackage[a4paper, total={7in, 9in}]{geometry}

\newcommand{\quotes}[1]{``#1''}

\newcommand{\SM}[1]{{Supplementary Note~{#1}}}
\newcommand{\FIG}[1]{{Fig.~\ref{#1}}}
\newcommand{\EQ}[1]{{Eq.~\ref{#1}}}

\begin{document}

	\title{Ranking nodes in bipartite systems with a non-linear iterative map}
	
	\author[1]{Andrea Mazzolini}
	\affil[1]{Laboratoire de physique de l’École normale supérieure (PSL University), CNRS, 75005 Paris, France}
	
	\author[2]{Michele Caselle}
	%\affiliation{Department of Physics and INFN, University of Turin, via P. Giuria 1, 10125 Turin, Italy}
	\author[2]{Matteo Osella}
	\affil[2]{Department of Physics and INFN, University of Turin, via P. Giuria 1, 10125 Turin, Italy}

	\maketitle

	\begin{abstract}
		
		Ranking nodes in networks according to a defined measure of importance is an extensively studied task, with applications in ecology, economic trade networks, and social networks.
		This paper introduces a method based on a non-linear iterative map to evaluate node  relevance in bipartite networks. 
		By tuning a single parameter $\gamma$, the method captures different concepts of node importance, including  established measures like degree centrality,  eigenvector centrality and the fitness-complexity ranking. 
		The algorithm's flexibility  allows for efficient ranking optimization tailored to specific tasks, outperforming state-of-the-art algorithms. 
		We apply this method to ecological mutualistic networks, where ranking quality can be  assessed by the extinction area - the rate at which the system collapses when species are removed in a certain order.
		The map with the optimal $\gamma$ value surpasses existing ranking methods on this task. Additionally, our method excels in evaluating nestedness, another crucial structural property of ecological systems, requiring specific node rankings.
		Finally, we explore theoretical aspects of the map, revealing a phase transition at a critical $\gamma$ dependent on the data structure that can be characterized  analytically for random networks.  
		Near the critical point, the map exhibits unique features and a distinctive   \quotes{triangular} packing pattern of the incidence matrix. 
		
	\end{abstract}

	\section{Introduction}

	How to quantify the \quotes{importance} of a node in a graph is an extensively studied question in network theory. 
	Several definitions of node centrality have been proposed, each exploiting different aspects of the network structure \cite{newman2018networks}.
	The simplest one is the well-known \textit{degree centrality}, in which the importance of a node is simply its degree \cite{freeman1978centrality}. The natural extensions of this idea are  centrality measures,  which take into account also the importance of the node's neighbors. The \textit{eigenvector centrality} \cite{bonacich1987power} and the \textit{Katz centrality} \cite{katz1953new} are two examples. 
	The popular \textit{PageRank centrality} \cite{brin1998anatomy} is also based on a similar concept.  In this case, the importance of a node is a linear function of the importance of its neighbors, but their contribution  is normalized by their out-degree.
	%This prevents peripheral nodes connected to a huge hub to have a large measure of centrality.
	While these centrality measures are the most relevant for this paper, there are plenty of other examples in network theory \cite{freeman1977set, kleinberg1999authoritative, newman2005measure, estrada2005subgraph, borgatti2006graph}.

	Bipartite networks are a particular class of networks in which links only connect nodes from two distinct sets. 
	This structure describes a vast number of complex systems, specifically component systems. In component systems, each realization is an assembly of basic building blocks, such as books composed of words or genomes composed of genes, which naturally translates into a bipartite network of connections between basic components and realizations~\cite{mazzolini2018statistics,mazzolini2018zipf,Mazzolini2018ssr,Lazzardi2023}. 
	This analogy has led to a fruitful exchange of methods between network theory and statistical data analysis in different domains~\cite{gerlach2018,valle2020,valle2022}.
	
	While extensions of node centrality measures to bipartite networks have been proposed,  the problem of node ranking in these ubiquitous structures is still an active area of research. 
	A possible naive approach is to project the links of the bipartite network on one of the two node sets, and then  directly use standard tools of network analysis on the resulting unipartite structure~\cite{watts1998collective, newman2001structure, cancho2001small, everett2016centrality}.  However, the projection clearly hinders a fundamental property of the network  with possible  relevant consequences on the analysis results~\cite{guillaume2004bipartite, guillaume2006bipartite, latapy2008basic, pavlopoulos2018bipartite}. 
	Few methods that explicitly keep  into account the bipartite structure  have been proposed, such as  extensions of the eigenvector centrality~\cite{hidalgo2009building, daugulis2012note},  or methods based on the statistics of  random walkers on the network~\cite{Coscia2014}.  
	Finally, a statistical approach based on a non-linear map, the so-called fitness-complexity map, was introduced in the context of economics,  and then generalized and applied in other fields~\cite{tacchella2012new,cristelli2013measuring,tacchella2018dynamical,
		cristelli2015heterogeneous, dominguez2015ranking,mariani2015measuring}
	
	The present paper proposes a general and flexible method to rank the importance of nodes in  bipartite networks based on a non-linear map inspired by the fitness-complexity concept.  
	The non-linearity of the map depends on a single parameter $\gamma$.  
	The parameter value sets what structural features are relevant for defining the  node importance and thus the final node ranking. Indeed, tuning $\gamma$ we can interpolate between the  eigenvector centrality, the degree centrality, and the original fitness-complexity map. 
	Depending on the system under analysis and the specific task, different choices of $\gamma$ can be optimal.

	Ecological networks provide an illustrative example in which different algorithms, and the resulting rankings, can be quantitatively compared~\cite{dominguez2015ranking,Mariani20191}. 
	In fact,  the so-called extinction area can be used to quantify how relevant a node ranking is in complex ecological networks.
	This quantity measures how fast the network collapses when nodes (e.g., species) are removed from the system in a specific order. Therefore,  the ranking of species maximizing the extinction area can inform preservation strategies and interventions in endangered ecosystems. 
	The fitness-complexity map was reported to outperform  previously existing algorithms when applied to this class of mutualistic systems \cite{dominguez2015ranking}.  
	We will show that our generalized non-linear map can provide even better results by choosing the appropriate value of $\gamma$.

	In the same context of ecological systems,  nestedness is a long-studied structural property~\cite{patterson1986nested, ulrich2009consumer,Mariani20191}.  
	A nested ecological system is essentially composed by  few \textit{generalist} species that tend to interact with almost all the other species,  and by several \textit{specialists} that preferentially interact with generalists,  rather than with other specialists. 
	The identification of nested structures and the generalist species at their core can help guide the design of conservation and preservation efforts, e.g., \cite{cutler1994nested, pocock2012robustness}. Therefore, in addition to the theoretical interest in understanding the evolution of ecological networks, accurate nestedness measures also have practical importance.
	Several popular  measures of nestedness require finding the ordering of rows and columns of the incidence matrix that maximizes its \quotes{triangular shape}. We will show that again our algorithm improves the performance of the fitness-complexity map, which, in turn, has been shown to perform better than previous state-of-the-art algorithms \cite{lin2018nestedness}.

	Having established the practical usefulness of this flexible non-linear map, we will focus on its general mathematical properties and on the role of the  parameter $\gamma$. 
	More specifically, we will show that the dynamical system defined by the map undergoes a  phase transition in its convergence properties for a critical value $\gamma_c$ that depends on  structural network properties.  We will explore in detail the close connection between the phase transition,  the maximization of the extinction area, and the re-ordering of the incidence matrix in a triangular form,  with all the ones  located in the upper-left corner and the bottom-right corner only composed by zeros. This  connection is also instrumental to define the parameter region where the value of $\gamma$ maximizing the extinction area can be found.

	\section{Methods}

	\subsection{Datasets}
	\label{sec:data}
	
	The ecological interaction matrices used in this work were downloaded  from Web of Life (http://www.web-of-life.es/), an open access database with a large number of published species interaction networks.
	The full list of matrices used in the {result section is  reported in the Supplementary Table 1.
	We primarily focus on large systems, i.e., those with a product of the number of rows and columns of at least 500, which are, in general, harder to treat with a full exploration of possible rankings or the application of computationally expensive genetic algorithms. 
	In particular,  the illustrative  example discussed in the result section is a  plant-pollinator interaction matrix based on the work of C. Robertson who collected in 1929 the list of  interactions between plants and insects in Carlinville, Illinoise.
	This dataset was validated and updated in ref.~\cite{marlin2001native}, extending the network to 1044 animal species visiting  456 plant species.

	\subsection{A general and flexible measure of node importance for bipartite systems}
	\label{sec:gen}
	
	\begin{figure*}[htbp]
		\centering
		\includegraphics[width=\textwidth]{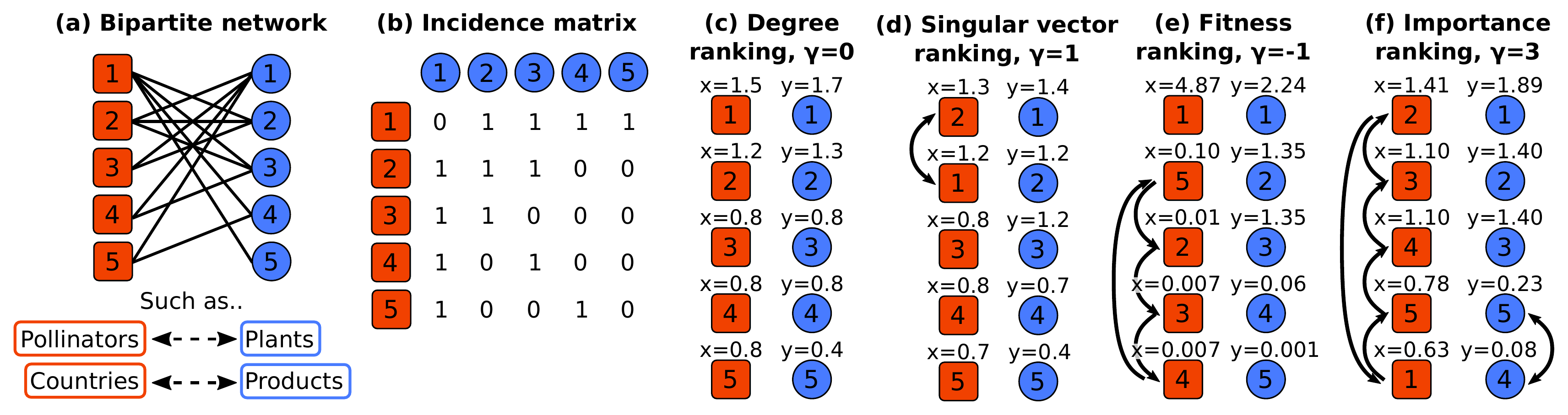}
		\caption{\textbf{Toy example of a bipartite system and its node ranking}.
			The network is represented as a non directed, unweighted, bipartite network, \textit{(a)}, where a set of squared nodes, e.g. countries or pollinator animals, interact with a disjoint set of circular nodes, e.g. products or plants.
			Such a network is associated with the incidence matrix \textit{(b)}, in which the element is equal to $1$ if there is a link between the two nodes.
			The panels from (c) to (f) show the scores and the rankings that \EQ{eq:gen_pn_map} provides for four values of $\gamma$ and are discussed in the main text. 
			The black arrows between nodes are changes in the ranking with respect to the degree ranking.}
		\label{fig:exe}
	\end{figure*}

	This section presents the non-linear map and how to use it to measure the node importance for bipartite networks.
	We consider unweighted undirected bipartite networks, with an $N \times M$ incidence matrix $A$, with $A_{ij} \in \{0,1\}$ and $A_{ij}=1$ if node $i$ is connected to node $j$ on the opposite node set (\FIG{fig:exe}a,b).
	We denote with $\mathbf{x}$ the scores of the nodes in the first node set of the network, and with $\mathbf{y}$ the scores relative to the second one.
	The scores are iteratively updated by the map, which starts from vectors of ones as initial conditions, i.e., $x_i^{(0)} = 1$ and $y_j^{(0)} = 1$, and is defined by the equations 
	\begin{equation}
		\left\lbrace
		\begin{aligned}
			& x_i^{(t)} = \frac{\tilde{x}_i^{(t)}}{\langle \tilde{\mathbf{x}}^{(t)} \rangle}, \; \text{ where } \; \tilde{x}_i^{(t)} = \sum_{j=1}^M A_{ij} \left( y_{j}^{(t-1)} \right)^\gamma ,
			\\
			& y_j^{(t)} = \frac{\tilde{y}_j^{(t)}}{\langle \tilde{\mathbf{y}}^{(t)} \rangle}, \; \text{ where } \; \tilde{y}_j^{(t)} = \sum_{i=1}^N A_{ij} \left(x_i^{(t-1)} \right)^\gamma.
		\end{aligned}
		\right.
		\label{eq:gen_pn_map}
	\end{equation}
	The angular brackets denote the vector average $\langle\mathbf{x}\rangle = \sum_{i=1}^N x_i / N$.
	The map, possibly, converges to stationary values, which represent  the importance measure of the nodes.
	In general, there is no guarantee that a stationary condition exists. However, we could define a ranking of the nodes from the map behavior at large $t$ for  all the empirical systems considered. 
	A detailed discussion about the map convergence is postponed to the section about the phase transition. 
	
	The numerical approximations  that have to be applied to reliably simulate the map in the presence of possible score  divergences, and a pseudocode are discussed in detail in the \SM{1}.
	The algorithm, and some of its applications,  can be downloaded from the public repository \url{https://github.com/amazzoli/xymap.git}.
	
	The map depends on the free parameter $\gamma$ which can be tuned depending on the idea of node importance suggested by the system in analysis. 
	In particular,  the generalized map recovers known measures of importance for  specific values of $\gamma$, as detailed in the following.

	\subsubsection{Degree centrality, $\gamma = 0$}
	
	The simplest case is the degree centrality, which is defined as the number of neighbors of a node, and can be obtained for $\gamma = 0$.
	In this case, the stationary solution is reached at the first map iteration, and the scores are simply the node degree divided by the average degree.
	\FIG{fig:exe}c shows the degree ranking of the toy example illustrated in \FIG{fig:exe}a.

	\subsubsection{Singular vector centrality, $\gamma=1$}
	
	For $\gamma=1$ the importance of a node is proportional to the importance of the connected nodes in the other node set. 
	In other words, a node is important if it has a lot of  important neighbors. The illustrative example in \FIG{fig:exe}d shows that this notion of importance leads to a different ranking with respect to  the case of $\gamma=0$ (i.e., degree centrality). In particular, the second squared node now surpasses the first one in the ranking because, even if it has less connections, it is connected to the most important circular node in the opposite side of the network. 
	
	This is the foundational idea of the eigenvector centrality in classical unipartite networks \cite{bonacich1987power}. According to this measure, the importance $c_i$ of a node $i$ is 
	\begin{equation*}
		c_i = \lambda^{-1} \sum_{j=1}^N B_{ij} c_j,
		\label{eq:eigen_centr}
	\end{equation*}
	where $B$ is the $N \times N$ adjacency matrix, which we assume to be symmetric and irreducible, i.e., the network is undirected and connected.
	The equation above, when written in matrix form,  is satisfied by each eigenvector of $B$ with corresponding eigenvalue $\lambda$: $B \mathbf{c} = \lambda \mathbf{c}$.
	To resolve the ambiguity of having multiple possible choices for the scores,  the scores are required to be all positive.
	This constrains the choice to the eigenvector corresponding to the largest eigenvalue, since the Perron-Frobenius theorem proves that it is the only eigenvector with all positive components ~\cite{hogben2013handbook}.
	Therefore, the leading eigenvector of $B$ defines the so-called eigenvector centrality.
	This definition  of score is arbitrary to within a multiplicative constant.
	However, one is usually interested in the node ranking, which is not affected by this degree of freedom.
	An important property of the eigenvector centrality is that it can be computed in a dynamical way, in the similar spirit of the map given by \EQ{eq:gen_pn_map}.
	Starting from a vector of ones as initial condition, the recursive relation $\mathbf{c}^{(t+1)} = \lambda^{-1} B \mathbf{c}^{(t)}$, converges to the leading eigenvector, in the limit of large $t$.
	
	This measure can be extended to bipartite networks \cite{daugulis2012note}.
	In this case, we can consider the rectangular incidence matrix $A$ and substitute the concept of eigenvectors with left/right singular vectors, and eigenvalues with singular values.
	\SM{2} shows the proof that the iteration of the \EQ{eq:gen_pn_map} for $\gamma=1$ leads to a vector of scores $\mathbf{x}$ which is proportional to the leading left singular vector $\mathbf{u}_1$ of $A$ (corresponding to the largest singular value $\sigma_1$), while $\mathbf{y}$ is proportional to the leading right singular vector $\mathbf{v}_1$:
	\begin{equation}
		\begin{aligned}
			\mathbf{x} & \propto \mathbf{u}_1
			\\
			\mathbf{y} & \propto \mathbf{v}_1
		\end{aligned}
		\hspace{0.5cm}\text{where}\hspace{0.5cm}
		\begin{aligned}
			A^T \mathbf{u}_1 = \sigma_1 \mathbf{v}_1
			\\
			A \mathbf{v}_1 = \sigma_1 \mathbf{u}_1
		\end{aligned}
		\label{eq:sing_vector_centrality}
	\end{equation}
	Again, the Perron-Frobenius theorem guarantees that the leading singular vectors are positive, and therefore consistent with the unique definition of a positive score.
	
	It is useful to describe one possible example of application of this score system: rating the desirability of products which have been used by a set of consumers \cite{daugulis2012note}.
	This can be described by a consumer-product bipartite system which has links whether a consumer has tried a product.
	Clearly, the more a product is consumed (high degree) the more important it is, but one can also speculate that an \quotes{experienced} consumer increases further the importance of the products that it has tried.
	The evaluation of consumer experience can be made by the amount of the tried products and if those tried products are considered very desirable.
	Another important example is the measure of economic complexity introduced in \cite{hidalgo2009building} (with a small difference in the normalization factor of the map).

	\subsubsection{Fitness-complexity map, $\gamma=-1$}
	
	The choice of the exponent $\gamma = -1$   recovers the fitness-complexity map introduced in the context of economics~\cite{tacchella2012new, cristelli2013measuring, cristelli2015heterogeneous}.
	The score $x_i$ corresponds to the fitness, while the score $y_j$ is the inverse of the complexity. Therefore, the fitness-complexity map is recovered through the substitution $\tilde{q}_j = \tilde{y}_j^{-1}$ for every $j$, and defining the complexity as $q_j = \tilde{q}_j / \langle \tilde{q} \rangle$. 
	
	In this regime,  the map attributes  large importance to nodes with  many connections, but, differently from the previous case of $\gamma=1$,  the connections  with low-importance neighbors have a larger weight.
	This effect is shown by the ranking in \FIG{fig:exe}e, where the squared node $5$ increases its position in the ranking with respect to the degree sorting, and surpasses the higher-degree node $2$. Indeed, node $5$   is connected to the circular node $4$, which has the second lowest importance, while the squared nodes $2$, $3$, and $4$, are not.
	
	The \textit{fitness-complexity} map was introduced to quantify the non-monetary competitiveness of a country on the basis of its exported products.
	The basic idea is that the fitness (i.e,   the $x$ score)  of a country is proportional to the sum of the complexities of its exported products, which in our notation is the sum of the inverse of the \quotes{simplicities}, given by the $y$ scores.
	The simplicity of a product is large if many countries can export that product, in particular if these countries have low fitness.
	On the contrary, a product is complex ( i.e., low $y$) if only few countries with high fitness can produce it.
	
	The range of application of the fitness-complexity map extends to other bipartite networks. 
	For example, it has been used to analyze the  contribution of different countries  to a set of  scientific topics ~\cite{cimini2014scientific}.
	%In this case the country-topic matrix has a 1 if a specific  country contributes significantly to a particular topic  using citations as a measure. The fitness-complexity map provides a ranking of scientific competitiveness of  nations.
	A different field of application are  mutualistic ecological networks.
	In the case of the interactions between plants and pollinators, the \quotes{fitness} of an animal pollinator seems to be significantly related to its ecological importance within the ecosystem, while the \quotes{complexity} of a plant quantifies its vulnerability to system perturbations \cite{dominguez2015ranking}.
	This ecological example will be our main illustrative application.

	\subsubsection{Behavior at large exponents}
	\label{large-exp}
	
	It is instructive to characterize  the map behavior for large exponents.
	In the case of $\gamma \gg 1$, the first score of \EQ{eq:gen_pn_map} can be approximated as:
	\begin{equation}
		x_i^{(t)} \propto \frac{\max_{j \in J(i)} \left( y_j^{(t-1)} \right)^\gamma} {\max_{j \in \{1, \ldots, N\}} \left( y_j^{(t-1)} \right)^\gamma},
		\label{eq:large_g}
	\end{equation}
	where the maximum at numerator comes from the approximation of $\tilde{x}^{(t)}$, and it is evaluated among the set  $J(i)$ of neighbors of $i$. 
	The maximum at denominator is instead the leading term of $\langle \tilde{x}^{(t)} \rangle$, and considers the largest y-score among all the nodes of the second node set.
	A similar expression can then be found for the update of the $y$ scores, whose values depend only on the maximal $x$ of their neighbors.
	This expression implies that only the nodes connected with the maximal y-score have non-negligible x-score and, therefore, will dominate the ranking.
	In this limit, the degree of the node does not enter explicitly in the computation.
	This can be seen in \FIG{fig:exe}f, which displays a ranking calculated for $\gamma = 3$.
	The squared node $1$, which is the most connected among the squared nodes, becomes the last one in the ranking, since it lacks the connection with the most important circular node, i.e., the first one.
	The opposite limit of $\gamma \rightarrow - \infty$ leads exactly to the same result of \EQ{eq:large_g}, but with a minimum instead of a maximum. Therefore, the $x$-score is  determined by the connection with the minimal $y$-score.

	\subsubsection{General intuition behind the map behavior}
	
	Putting all these considerations together one can conclude that the absolute value of $\gamma$ sets the balance between two properties in determining the importance of a node.
	The first is the degree of the node which, for $\gamma = 0$, is the only relevant feature.
	The second property is the connection with important nodes in the other node set in the limit of large $|\gamma|$, only the maximum (or minimum) score among the connected neighbors contributes to the importance.
	%In this second limit the degree does not enter directly in the computation. 
	%However,  it is still playing an indirect role, since having a lot of neighbors can increase the chance of being connected with these extreme-score nodes.
	The sign of $\gamma$ determines if the importance is given by having high-score neighbors (positive $\gamma$) or low-score ones (negative $\gamma$).

	\section{Results}
	\label{sec:results}

	\subsection{Looking for the specie ranking which maximizes the extinction area}
	\label{sec:ext_area}

	We can now evaluate the ranking generated by the map on ecological systems of mutualistic interactions between species, such as plants and pollinators. 
	We  use the extinction area to  quantitatively evaluate the goodness of the rankings~\cite{dominguez2015ranking, allesina2009googling}. The computation of this quantity is illustrated in \FIG{fig:ext_area}a-c.
	Given an animal  ranking (e.g., $(A_1, A_2, A_3)$), we can first remove the first animal ($A_1$) and all its links. As a consequence, a certain number of plants remain without links and thus goes extinct ($F_1$ in the example).  
	
	By progressively  removing all  the animals according to the ranking, we can calculate the fraction of extinct plants as a function of the fraction of removed animals (\FIG{fig:ext_area}c). 
	The extinction area is defined as the integral of this curve, and quantifies the speed of the ecosystem collapse for the given  ranking.

	\begin{figure}[htbp]
		\centering
		\includegraphics[width=0.45\textwidth]{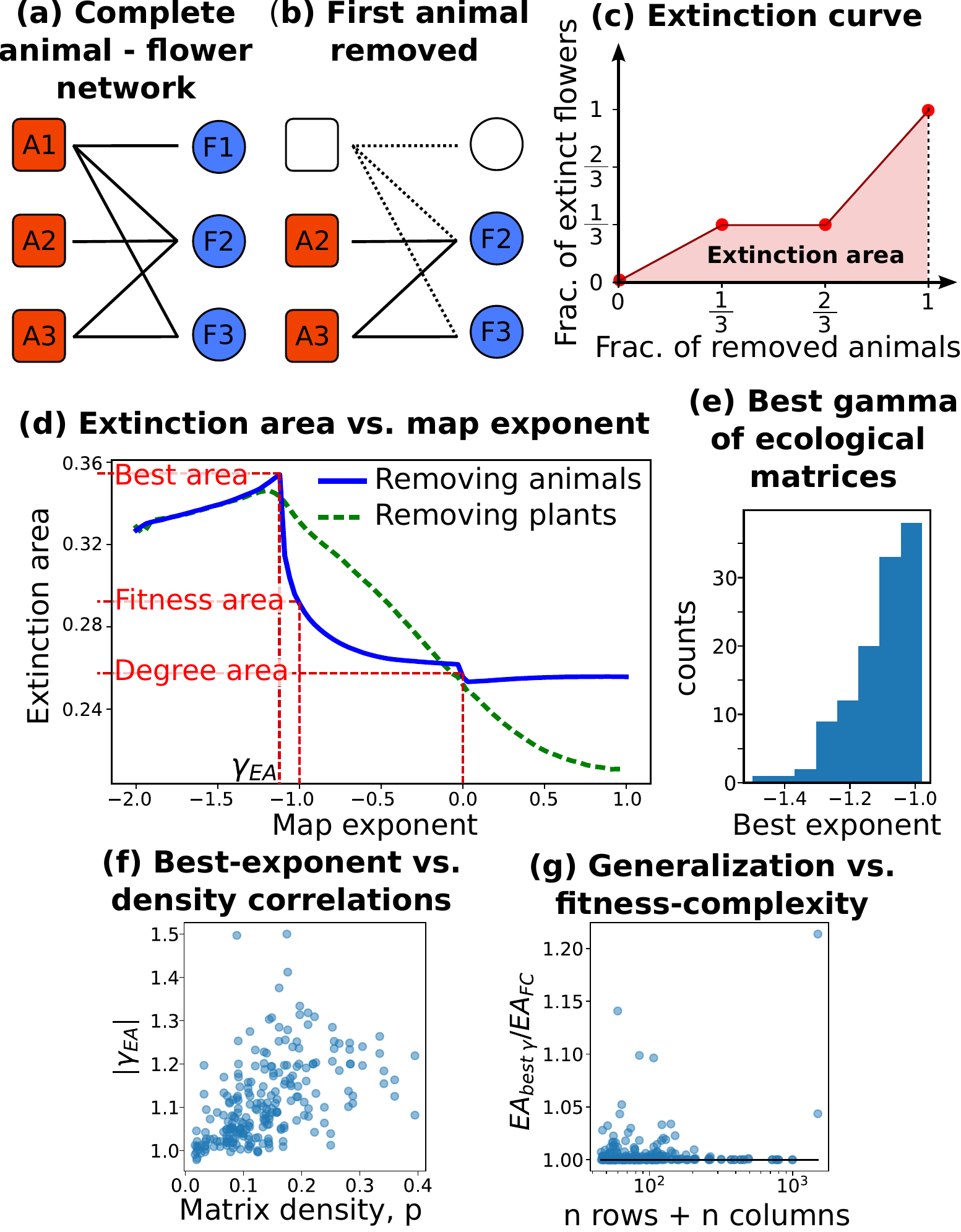}
		\caption{\textbf{Computing the extinction area in ecological networks.}
			The panels \textit{(a)}, \textit{(b)} and \textit{(c)} show a toy example of the extinction area computation. 
			Starting from the complete network \textit{(a)}, animals (matrix rows) are removed according to a certain ranking.
			In \textit{(b)} the animal $A_1$ has been deleted leading to the extinction of the plant $F_1$ which remains without links.
			The extinction area is the integral of the curve \textit{(c)} which shows the fraction of extinct plants at a given fraction of removed animals.
			The panel \textit{(d)} is the extinction area computed at different map exponents $\gamma$ of the mutualistic system Robertson 1929.
			The blue continuous line correspond to animal removal, where the animal ranking is provided by the $x$-score at the given $\gamma$.
			The green dash-dot line is computed using the opposite procedure: the plants are progressively removed following the $y$ ranking.
			The panel \textit{(e)} shows the number of matrices whose extinction area by animal removal is maximized at given exponents, $\gamma_{EA}$, among the  $116$ mutualistic networks of the Web-of-Life database, see \ref{sec:data} and table S1, having size (product between the number of rows and columns) larger that $500$. 
			Panel \textit{(f)} shows the correlation between matrix density and map exponent that maximizes  the extinction area - Spearman correlation $\rho = 0.61$, $\text{p-val}=10^{-24}$. 
			Panel \textit{(g)} reports the  ratio between the  extinction areas obtained with the fitness-complexity map with  our generalization. We consider both rows and columns removal procedures.
		}
		\label{fig:ext_area}
	\end{figure}

	A previous work \cite{dominguez2015ranking} demonstrated that the fitness-complexity map, that we recover for $\gamma=-1$, provides an animal ordering that leads to a larger extinction area than other algorithms.
	\FIG{fig:ext_area}d shows that, in a specific case, our map can reach even larger values of extinction area for other values of $\gamma$.
	In particular, there is a maximum at $\gamma_{EA} \simeq -1.12$.
	The improvement with respect to the fitness-complexity map is even larger than the improvement that the fitness-complexity map has compared to the simple degree ranking ($\gamma=0$).
	
	The value of $\gamma$ that maximizes the area, $\gamma_{EA}$, is dataset dependent. 
	We notice that in all the cases we studied  $\gamma_{EA}$ was strictly smaller than -1 and typically within the range $\gamma_{EA} \sim [-1.4,-1]$. 
	We report in \FIG{fig:ext_area}e the distribution of the best parameters for all the mutualistic networks discussed in \quotes{Dataset} section.
	We shall further discuss the issue of finding the best gamma and, in particular, the origin of the $\gamma_{EA} < -1$ found in result section about the phase transition.
	
	Given the variability in the exponents that maximize the extinction area across different ecosystems, we can test whether simple statistical features of the interaction matrices correlate with these values.
	The most significant (anti)correlation we observe is with matrix density, i.e., the fraction $p$ of actual links among all possible ones.
	In fact, \FIG{fig:ext_area}f suggests that very sparse matrices are optimally ranked by exponents close to one, whereas denser matrices require smaller exponents.
	The corresponding analysis for uniform random matrices, reported in Supplementary Figure 8, confirms that, on average, smaller exponents are needed for denser matrices.
	For random matrices, we also observe a clear dependence on system size (i.e., the number of species) at a given interaction density.
	The possible reasons behind these dependencies are further discussed later.

	For several systems, the exponent $\gamma_{EA}$  that maximizes the extinction area is close to $-1$ (\FIG{fig:ext_area}e), where our map becomes  equivalent to the  fitness-complexity approach. Indeed, for these interaction networks, our map rankings provide little to no improvement in the extinction area. However,   \FIG{fig:ext_area}g highlights several notable exceptions where our map leads to significantly larger extinction areas —by up to 20\% — as observed for the ecological system used for panel d.
	In these cases, the species ranking obtained with  $\gamma_{EA}$ can be significantly  different from the ranking corresponding to $\gamma=-1$. Supplementary Figure 9 analyzes the ranking variations using $\gamma_{EA}$ or  $\gamma=-1$ for the specific network analyzed in panel d.
	The rank is well conserved only for the lowest-ranked species. However, in many cases, species shift by as many as 100 positions (out of a total of 1044 species), with some changing by up to 500 positions. The top two species remain the same in both rankings, but the top 10 differs dramatically. Species ranked in the top 10 of the fitness-complexity ranking drop to positions 200 or 300 in the best-gamma ranking, and vice versa.
	Consistently with the interpretation discussed in the method section, since  $\gamma_{EA}$  is relatively  smaller and far from   $\gamma=0$, species degree plays a less significant role in determining the ranking. The color pattern of Supplementary Figure 9 shows that  the increased importance of \quotes{low-importance} connected species in the other set in the  ranking.
	In this example, using the fitness complexity map  all top-10 species have a large degree (around 100), while in the $\gamma_{EA}$ ranking only two species have large degree, 5 have an  intermediate degree ($\sim 50$) and three have a small degree ($\sim 10$).

	One further question is how close the obtained extinction area is to the global maximum.
	We can explore a large population of rankings by employing  a genetic algorithm that searches for the best extinction area.
	\SM{3} shows that, for different ensembles of randomly generated matrices, the average best extinction area found by our map always coincides or is better than the genetic-algorithm result.
	It is worth noticing that the iterative map is typically 100 times faster than a genetic algorithm, as also discussed in \SM{3}.
	In the same section,  we analyze the extinction area obtained by different  algorithms when applied to the small ecological matrices of the Web of Life database, for which we can run a genetic algorithm in a reasonable amount of time.
	Also in this case our map  find the same extinction area of the genetic algorithm, but for those very small ecosystems the ranking difference with respect to  the fitness-complexity map is negligible.

	Finally, we investigated possible differences in progressively removing rows or columns. As panel A shows, there can be differences in the EA value at fixed exponent, but the maximum is roughly at the same value.
	Supplementary Figure 10 indeed shows that the optimization procedures by removing rows or  columns lead to almost the same maximum for each matrix.
	There is more variability in the exponents at which the maximum is attained using the two alternative  removal procedures. The exponents often do not coincide precisely, but are however strongly correlated  (Supplementary Figure 10).

	\subsection{Looking for the species ranking that maximizes the matrix nestedness}
	\label{sec:nest}
	
	\begin{figure}
		\centering
		\includegraphics[width=0.47\textwidth]{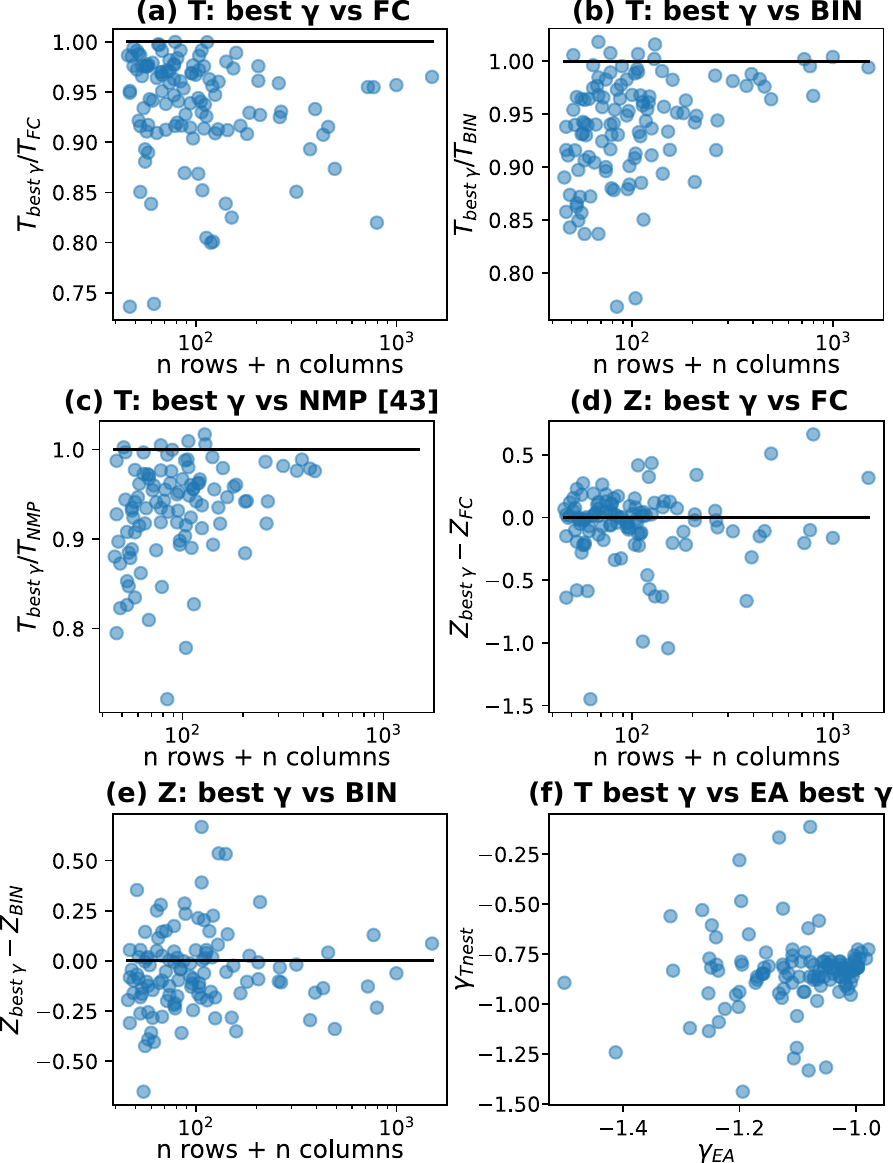}
		\caption{\textbf{Nestedness in ecological networks.}
			\textit{(a)}: ratio of the nestedness temperature between the fitness-complexity map, $\gamma=-1$, $T_{FC}$, and the best temperature found by our algorithm, $T_{best \;\gamma}$.
			On the x-axis we show the matrix size as the sum of the number of rows and columns.
			Each point is an ecological matrix of section \quotes{Dataset}.
			\textit{(b)}: temperature ratio using the ordering of BINMATNEST, from the R package vegan, nestedtemp method.
			\textit{(c)}: temperature ratio using the ordering obtained from the procedure discussed in \cite{mariani2024ranking}. 
			We used the algorithm of the repository associated to the paper with an hyperparameter $\beta=50$. Notice that the algorithm failed for around 10\% of the matrices, for which we do not report any point.
			\textit{(d)}: difference in the temperature Z-score for best-gamma ranking and the fitness complexity ranking. The underlying null model preserves on average the degree of rows and columns.
			\textit{(e)}:  difference in the temperature Z-score for best-gamma ranking and the BINMATNEST ranking.
			\textit{(f)}: for each matrix we plot the value of $\gamma$ that maximizes the extinction area, y-axis, and the one that maximized the nestedness temperature, x-axis.
		}
		\label{fig:nest}
	\end{figure}

	The classical way to measure the nestedness of a network depends on the reordering of rows and columns \cite{atmar1993measure}.
	Given a row and column order, one can compute a nestedness \textit{temperature} and the final temperature of the network is the minimal one among all the possible orders.
	Therefore,  this measure relies on algorithms that have to explore the factorially growing space of possible rankings of rows and columns.
	
	The fitness-complexity map has proven to be highly effective for this purpose, identifying rankings that, in several cases, outperform state-of-the-art search algorithms such as BINMATNEST \cite{rodriguez2006new,lin2018nestedness}.
	This task can be tackled with our map  by varying the parameter $\gamma$ and choosing the value  minimizing the nestedness temperature.
	We adopted the temperature definition of ref.~\cite{rodriguez2006new} and \cite{lin2018nestedness}, whose implementation can be found in the repository associated to the manuscript \url{https://github.com/amazzoli/xymap.git}.
	The temperatures that we find are typically better than the ones at $\gamma=-1$ (\FIG{fig:nest}a).
	In particular, we find lower temperatures also for larger matrices, where the fitness-complexity map was performing slightly worse than the BINMATNEST, as reported in Fig.~3 of reference \cite{lin2018nestedness}.
	Furthermore, our map, with an optimized gamma value, systematically results in a lower temperature compared to BINMATNEST   (\FIG{fig:nest}b) or a more recent algorithm that addresses the Nestedness Maximization Problem (NMP) as an optimization problem using statistical physics techniques~\cite{mariani2024ranking} (\FIG{fig:nest}c).

	An absolute value of measured nestedness does not have a particular meaning per se. In ecology, one can compare the nestedness of different ecological communities or the empirical value against a suitable null model to determine its statistical significance~\cite{mariani2019nestedness}.
	The Z-score is a simple measure that can be used to assess significance of the difference in nestedness  temperature between a real network and the average over an ensemble of random networks,  expressed in units of standard deviations.
	We consider a random null model that conserves the degree of rows and columns on average and compute the average temperature and its standard deviation over an ensemble of 30 random matrices for each of the 116 empirical ecological networks. We can then compare the Z-scores obtained by measuring the temperature with our method,  with the fitness-complexity map (\FIG{fig:nest}d)   and with  BINMATNEST  (\FIG{fig:nest}e).
	The variations in the computed  Z-scores can go up to $0.5$ units of standard deviations.
	Despite not being a drastic change, it can still have consequences in specific cases. For example,  the network labeled as \quotes{M PL 046} in Web of Life database, move from  a $Z = -1.6$ computed with BINMATNEST to a $Z = -2$, possibly crossing a significance threshold.

	The best parameters for minimizing the nestedness temperature, $\gamma_{nest}$, take values in the range $[-1.5, 0]$.
	\FIG{fig:nest}f shows the scatter plot of this value and the one that maximizes the extinction area for the same matrix.
	One would naively expect that these two tasks are somehow similar, since they both try to \quotes{triangularize} the incidence matrix.
	However, the exponents for the two tasks take values in different regions of the parameter space, i.e.,  before and after $\gamma=-1$. Moreover, the values are not correlated (Spearman $\rho = 0.01$, $\text{p-val} = 0.9$).
	
	Finally, an alternative metric of nestedness is often used in the ecological literature with the acronym NODF, which stands for Nestedness metric based on Overlap and Decreasing Fill \cite{almeida2008consistent}.  
	While this measure still depends on the row and column order, its dependency is much weaker.
	In particular, our algorithm cannot find better rankings to maximize this metric than the simple degree ordering, which our map recovers for $\gamma = 0$ (Supplementary Figure 11). The  fitness-complexity map ($\gamma = -1$) and the two other methods tested in \FIG{fig:nest}b,c even lead to a slight decrease in this particular metric  (Supplementary Figure 11).

	\subsection{A phase transition suggests the value of $\gamma$ maximizing the extinction area}
	\label{sec:pht}

	\begin{figure*}[ht]
		\centering
		\includegraphics[width=0.95\textwidth]{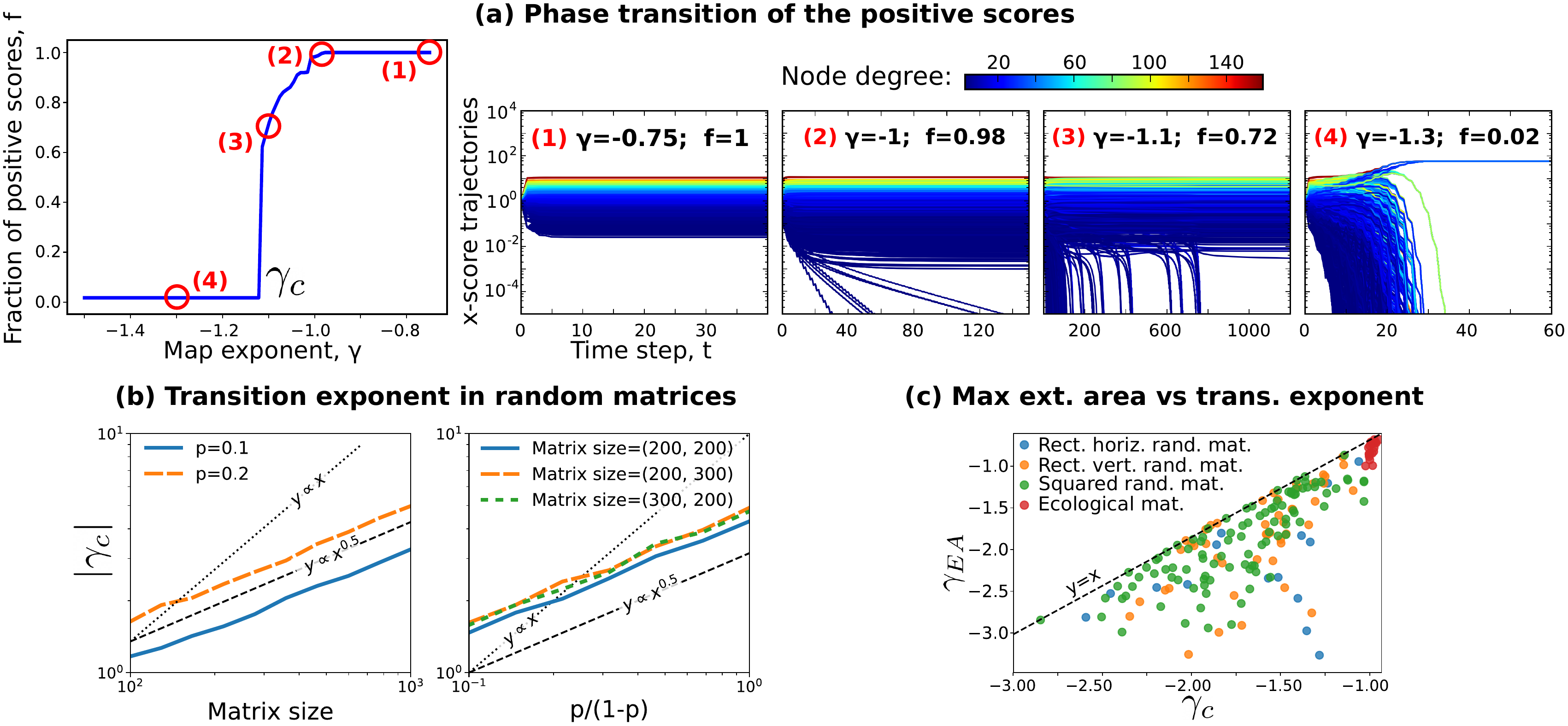}
		\caption{\textbf{Map phase transition.}
			The panel \textit{(a)} shows the transition of the fraction of scores which converge to positive values of the plant-pollinator matrix Robertson 1929 (see Sec. \quotes{Datasets}).
			The four red circles above the blue line of the plot on the left are associated to the four plots on the right, which show the $x$-score trajectories for different exponents. The line colors refer to the node degree.
			The panel \textit{(b)} displays the absolute value of the transition exponent, $\gamma_c$, for the $x$-score as a function of the size and the density in uniform random matrices.
			Different line colors and styles represent different matrix densities on the left and different matrix sizes on the right.
			Each point is computed as an average over $5$ matrices generated with the same parameters.
			The panel \textit{(c)} shows the scatter plot of the exponent which maximizes the extinction area, $\gamma_{EA}$ and the exponent of the phase transition, $\gamma_c$, for uniform random matrices and the $20$ largest ecological matrices (red points).
			The other three colors are different shapes of uniform random matrices: horizontal $N=s/1.5, M=s*1.5$ in blue, vertical $N=s*1.5, M=s/1.5$ in orange, and squared $N=M=s$  in green.
			The considered matrices have been generated with different sizes $s$, between $50$ and $200$, and different densities between $0.2$ and $0.4$.}
		\label{fig:pht}
	\end{figure*}  
	
	This section examines on the stationary behavior of the general map described by \EQ{eq:gen_pn_map} for different values of  $\gamma$. 
	Previous work~\cite{pugliese2014convergence, wu2016mathematics} provided some analytical insights on map  convergence properties as a function of the input matrix for the specific case of $\gamma=-1$ and for a generalization of the fitness-complexity map similar, but not equivalent, to our map (its comparison with our map is postponed to the discussion).
	
	As the number of iterations of the map increases, the scores $x_i$ and $y_j$ tend to an asymptotic behavior which strongly depends on the value of $\gamma$. For special choices of the adjacency matrix and of $\gamma$, this limiting behavior may even  be an oscillatory type. However,  the generic situation is that the scores tend to well defined and  finite fixed points. 
	Note that the divergence to infinite values is not allowed since the trajectories are normalized at each time step, limiting the maximum value of the score to the number of nodes in the corresponding layer.  
	We observe  that the fixed points are all strictly positive 
	for  positive parameters ($\gamma > 0$),  as well as for negative and small parameters, i.e.,   $\gamma < 0$ and $|\gamma| \ll 1$. An  example is reported in  \FIG{fig:pht}a1. On the other hand,  for small enough values of $\gamma$,  some of the scores can tend to zero as in the examples of  \FIG{fig:pht}a2-4.
	This latter phenomenology could create an ambiguity in the ranking definition. 
	However, not all the nodes' scores tend to zero with the same speed, and we can thus  rank first the nodes that have a slower score decay, as better explained in \SM{1}.
	
	Interestingly, how  the fraction of zero-scores changes as a function of $\gamma$ strongly resembles a phase transition. 
	This is displayed in \FIG{fig:pht}a for the particular case of the Robertson 1929 matrix.
	For exponents close to zero, the score is proportional to the node degree, implying that all the fixed points are positive: $f = 1$.
	Also when the exponent is decreased from zero to values near $-1$, for instance at $\gamma = -0.75$ in \FIG{fig:pht}a1, all the scores reach positive values. 
	Approaching $\gamma=-1$, a small fraction of trajectories begins decaying to zero, as in \FIG{fig:pht}a2.
	Moving from $-1$ towards a \quotes{critical} exponent, the fraction of positive trajectories decreases, as shown in \FIG{fig:pht}a3, for $\gamma = -1.1$.
	Finally, after a discontinuous transition, there is a "condensation phenomenon" with most trajectories converging to zero with few exceptions,  as in \FIG{fig:pht}a4 for $\gamma = -1.3$.
	As it typically happens in phase-transition phenomena,  the time required for  convergence increases approaching the transition.
	This abrupt transition from a macroscopic number of positive scores to a fraction of order $1/N$ is present across all the different empirical cases  we have considered (see \SM{4}).
	
	Moreover, the behavior of the order parameter as a function of $\gamma$ in the case of random interaction matrices $A_{i,j}$ shows that the  jump becomes steeper  with the matrix size. This is  exactly the expected phenomenology for a phase transition in the thermodynamic limit, i.e.,  in the large matrix size limit. This allows to define a critical value $\gamma_c$ which in the thermodynamic limit separates the $f\sim 1$ phase (for $\gamma> \gamma_c$) from the $f \sim 0$ phase (for $\gamma< \gamma_c$). In the empirical cases,  this critical value is rounded out due to the finite size of the interaction matrix, as it can be seen in \FIG{fig:pht}. However, despite finite size effects,  the critical $\gamma$ region   is  narrow and the transition in the order parameter is still evident.
	
	Even if the underlying dynamics is different, this phase transition strongly resembles the condensation phenomena discussed for instance in \cite{godreche2019condensation} or in \cite{Bianconi2001}. The score of one (or few) node takes a macroscopic value, while the other scores  are pushed to zero due to the normalization constraint.  For this reason,  we  denote the $f=0$ phase for $\gamma<\gamma_c$ as the \quotes{condensed phase} and the $\gamma_c$ as the \quotes{condensation point}.
	
	Unfortunately,  there is no obvious way to predict the value of $\gamma_c$ for a given empirical interaction matrix.
	However, the calculation of \SM{5} derives an upper bound in the limit of large matrices:
	\begin{equation}
		\gamma_c \le -1
		\label{eq:gamma_c_bound}
	\end{equation}
	
	While we have begun analyzing the relationship between the convergence transition and various properties of the input networks, there remains significant room for further investigation.
	For the fitness-complexity map and for a different generalization, a  previous work~\cite{pugliese2014convergence} proposed an ansatz for predicting the fraction of scores converging to positive values depending on the final reordering of the incidence-matrix rows and columns. Testing this intuition and its implications in the context  of our  map is a possible research direction.  
	An alternative approach is to simplify the class of input matrices in the analysis, as done in ref.~\cite{wu2016mathematics} for studying fitness-complexity convergence using perfectly nested matrices. In a similar vein, in the next section, we consider uniform random matrices, for which we derive a scaling relationship between the transition exponent and certain matrix properties.

	\subsubsection{Phase transition in uniform random matrices}
	
	In the case of uniform random matrices, the scaling of the critical map-exponent with the parameters of the incidence matrix can be understood with the following computation.
	Let us choose the matrix elements to be Bernoulli random variables, such that $A_{ij} = 1$ with probability $p$ and $A_{ij} = 0$ with $1-p$, where the matrix has size $(N,M)$.
	The map starts from initial conditions of ones, $x_i^{(0)} = 1$ and $y_j^{(0)} = 1$, and, after one step, it leads to sums of independent and identically distributed random variables: $\tilde{x}^{(1)}_i = \sum_{j=1}^{M} A_{ij}$.
	Using the central limit theorem one can obtain the following expressions:
	\begin{equation} \label{eq:step1}
		\begin{cases}
			{x}^{(1)}_i = 1 + \sqrt{\frac{1-p}{M p}} \eta \\
			{y}^{(1)}_j = 1 + \sqrt{\frac{1-p}{N p}} \eta,
		\end{cases}
	\end{equation}
	where $\eta$ is the standard normal random variable ($\langle \eta \rangle = 0$ and $\text{Var}(\eta) = 1$).
	At the next step, the non-normalized score reads:
	\begin{equation*}
		\tilde{x}^{(2)}_i = \sum_i^{M} A_{ij} \left( 1 + \sqrt{\frac{1-p}{N p}} \eta \right)^{\gamma}.
	\end{equation*}
	If $|\gamma|$ is small,  the score can be approximated at first order with the following small quantity: $|\gamma| \sqrt{(1-p)/(\min({N,M}) p)} \ll 1$.
	This leads to a summation of variables, both for $x^{(2)}$ and $y^{(2)}$, which have the same first and second moment of the Bernoulli matrix element.
	As a consequence, by applying again the central limit theorem, the score at the second step has the same distribution of the score at the first step (\EQ{eq:step1}), implying that the stationary solution has been reached. 
	The scores follow a Gaussian distribution, and are all positive. Thus for these values of $\gamma$ the system is in the $f=1$ phase.
	We know that for large enough negative values of $\gamma$ this solution is certainly unstable and the system is in the condensed phase.
	If we assume that the onset of the condensed phase coincides with the values of $\gamma$ for which the above approximation is not valid any more, i.e., $|\gamma| \sqrt{(1-p)/(\min({N,M}) p)} = O(1)$, we can derive the following scaling relationships for the critical exponent:
	\begin{equation}
		\gamma_c \sim - \sqrt{\frac{\min({N,M})p}{1-p}}.
		\label{eq:gamma_c_unif}
	\end{equation}
	Those predictions are verified in \FIG{fig:pht}b.

	\subsubsection{The connection between condensation and the maximal extinction area }
	
	The condensation of a few scores to positive values seems to be related to the good performance in finding the extinction area.
	This is shown by \FIG{fig:pht}c, where the transition exponent, computed according to the procedure described in \SM{4}, is plotted as a function of the exponent that maximizes the extinction area $\gamma_{EA}$, which is always located in the condensation phase. Combining this empirical observation with  \EQ{eq:gamma_c_bound}, we can derive the following bound:
	\begin{equation}
		\gamma_{EA} \le \gamma_c \le - 1, 
		\label{eq:ea_bound}
	\end{equation}
	which can explain the distribution of exponents maximizing the extinction area of different ecosystems in \FIG{fig:ext_area}e.
	Moreover, the bound induces a correlation between the two exponents, leading to similar dependencies on matrix size and density, at least for uniform random matrices.
	Supplementary Figure 8 illustrates the scaling behavior of the exponent that maximizes the extinction area with respect to matrix size and density. Indeed, these scalings resemble those of the critical exponent in~\FIG{fig:pht}b. 
	However, this connection between the two exponents does not fully account for certain  patterns we observed in empirical  ecological networks. For example, \FIG{fig:ext_area}f reports  a negative correlation between $\gamma_{EA}$ and matrix density,  whereas $\gamma_{c}$ shows no significant correlation with density ($\rho=0.06$, p-value$=0.5$).
	This suggests the presence of more complex relations in presence of structured non-random matrices between the phase transition, the matrix properties, and the system stability captured by the maximal extinction area.

	\subsection{The condensation implies a triangular shape of the incidence matrix}
	\label{sec:pack}

	\begin{figure*}[ht]
		\centering
		\includegraphics[width=\textwidth]{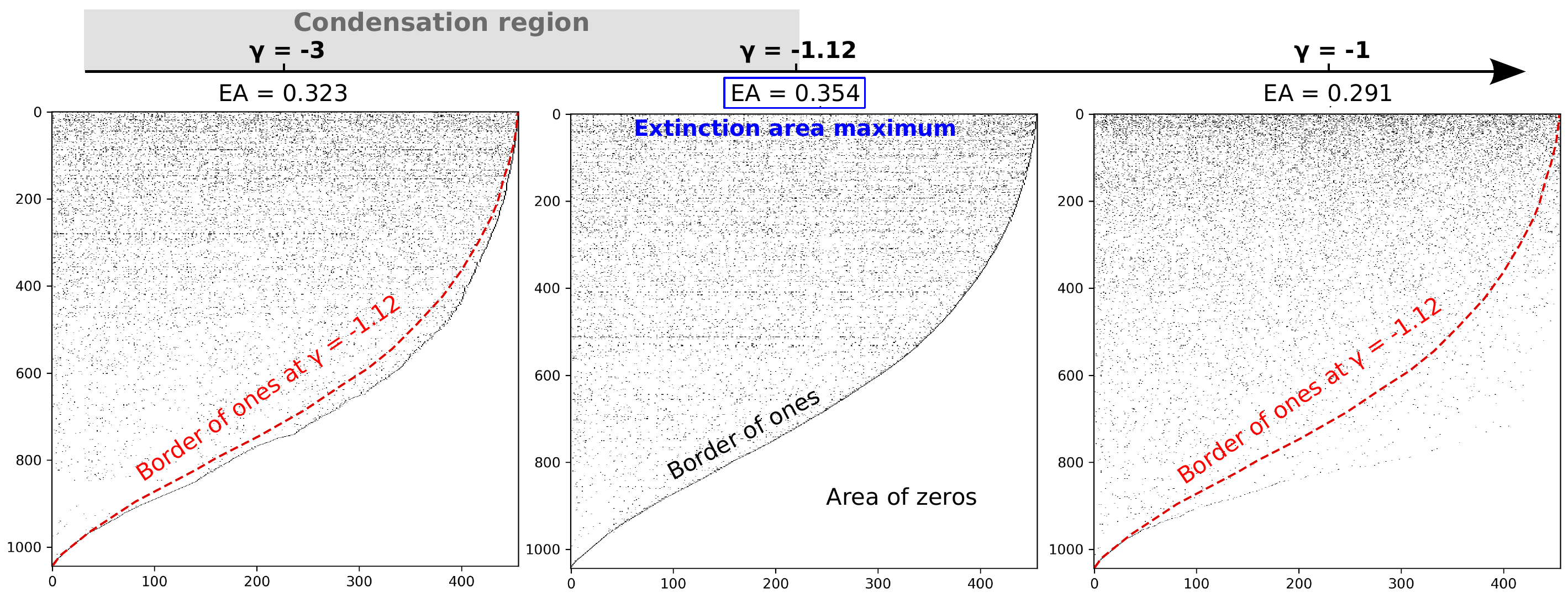}
		\caption{\textbf{Incidence matrix reordered with the map ranking.}
			Examples of matrices with rows and columns sorted according to the $x$ and $y$ ranking for three different map exponents, $\gamma = -3, -1.12, -1$, of the same mutualistic system (Robertson 1292 matrix).
			A black dot corresponds to the entry $1$, while a white dot to the entry $0$.
			$\gamma \simeq -1.12$ is the value which maximizes the extinction area, as shown in \FIG{fig:ext_area}d.
			The exponents less or equal than $-1.12$ are in the condensation phase, see \FIG{fig:pht}a, and present the typical border of ones that separates a right-bottom area of zeros from the rest of the matrix.}
		\label{fig:pack}
	\end{figure*}

	The shape of the incidence matrix, when ordered according to the map ranking,  shows a clear geometric pattern in the condensation phase.
	\FIG{fig:pack} shows the matrix corresponding to an ecological example (Robertson 1929), with the ones represented in black and the zero entries in white.
	Three different rankings are displayed corresponding to three different value of the exponent $\gamma$.
	In the condensed phase, a large area of contiguous zeros appears in  the bottom right corner. The area is well separated by the top left area by a continuous border of ones.
	The calculation of \SM{5} shows how both these features can be derived from the map (\EQ{eq:gen_pn_map}) with the only assumptions of having the system in the $f=0$ condensed phase, and the solution to be stable.
	
	There is a connection between this packing phenomenology and the extinction area. \SM{6} proves that the extinction area obtained by removing the rows is a linear function of the area of consecutive zeros from the matrix bottom.
	Specifically, if we define $z_j$ as the number of consecutive zeros from the matrix bottom of the column $j$, and $Z = \sum_1^M z_j$ the total area of zeros, the extinction area obtained removing rows, $E$, reads:
	\begin{equation}
		E = \frac{1}{N} \left( \frac{Z}{M} + 1\right) ,
		\label{eq:zeros_ext_area}
	\end{equation}
	where $N$ and $M$ are the number of rows and columns respectively.
	A similar relation can be derived also for the transposed quantities. 
	In other words, the extinction area obtained by removing columns (removing nodes from the second set and counting how many nodes of the first one get extinct) is linearly dependent on the area of zeros from the matrix right.
	These relations connect the extinction area with the visual triangular pattern of the adjacency matrix,  and we expect that the ordering with the largest area of zeros (second matrix of \FIG{fig:pack}) would also have the largest extinction area.

	The results of the last two sections suggest that the behavior of the map in the condensed phase drives the ranking of nodes in a way that it \quotes{triangularizes} the adjacency matrix, leading to a large area of zeros of the bottom right corner, which also  implies a large (almost maximal) extinction area. 
	It is important to stress that what we are able to prove analytically is the correspondence between the area of zeros in the right lower part of the matrix and the extinction area. Moreover, we can prove that in the condensation phase (i.e., for large enough negative values of $\gamma$), the matrix takes a triangular shape with a continuous border of ones. 
	However, we cannot prove that close to the transition the map provide an optimal packing of zeros corresponding to the higher extinction area. 
	As a general empirical fact, in all the studied cases the best extinction area is always reached in the condensation phase and close to the transition point, and therefore for $\gamma < -1$.

	\section{Discussion}

	Bipartite networks are the  natural description for  various  complex systems in   technological \cite{pang2013universal, mazzolini2018zipf}, social \cite{vasques2020transitivity, koskinen2012modelling}, economical \cite{hidalgo2009building, cristelli2013measuring}, linguistic \cite{gerlach2018network},  and biological \cite{corel2018bipartite, mazzolini2018statistics,valle2024exploring} domains. 
	We  introduce  a tool to rank nodes in bipartite networks according to different concepts of importance.  The scores used for ranking are defined through a  non-linear map connecting the two sets of nodes of the  network.   The map depends on the network incidence matrix and on an exponent $\gamma$ which dictates the non-linearity,  and implicitly defines  node importance.  
	By selecting $\gamma$,  users can tailor the analysis to the specific task. For example, the sign of $\gamma$ defines if a node is important when connected to high-score neighbors or low-score neighbors. We have shown how specific $\gamma$ values ($0$, $\pm 1$, $\pm \infty$) correspond to well known definitions of node importance.
	The interpretation of  other values of $\gamma$ is not trivial,  but qualitatively the $\gamma$ value informs about the importance attributed to the scores of connected nodes.
	More generally, given a metric to  measure how good  a ranking is,  we can look for the $\gamma$ value that maximizes it. For instance, we examined the problem of identifying the maximal extinction area in ecological interaction matrices, thus ranking the species according to their importance in the ecosystem stability. Analogously, we analyzed when our map can identify species rankings that minimize the nestedness temperature.
	These two tasks are central for understanding the structure of complex ecosystems and to inform preservation strategies by identifying  core species for the system stability.
	These tasks have been previously  effectively addressed using the  fitness-complexity map,  which correspond to   $\gamma=-1$ in our framework  \cite{dominguez2015ranking, lin2018nestedness}. Our results indicate that the optimal  $\gamma$ values   range from   $[-1.4,-1]$ for the extinction area,  and from $[-1, 0]$ for nestedness. 
	The optimal values vary based on the specific task and dataset, making them challenging to predict in advance. However, theory can help in limiting the range of parameters that should be explored. For example, we  could analytically set the upper bound at $\gamma_{EA} = -1$ for the extinction area,  \EQ{eq:ea_bound}.
	Our theoretical argument is based on the presence of a dynamical phase transition for large enough negative values of $\gamma$. This transition separates the score space into two phases. For $\gamma > \gamma_c$ all  scores remain  finite and positive, whereas for $\gamma <\gamma_c$ a \quotes{condensed} phase emerges  where only one or few scores dominate,    pushing the others towards zero through 
	normalization.
	We first established an upper bound for this transition as $\gamma_c \le 1$,  and then we postulated (and check numerically, \FIG{fig:pht}) that the extinction area is maximized in the condensed phase, $\gamma_{EA} \le \gamma_c$.
	This picture is coherent with the visual inspection of the incidence matrix ordered according to the ranking in the condensed phase.
	We proved that the extinction area is proportional to the area of consecutive zeros from the bottom-right of the matrix and, indeed, the matrix takes a clear triangular shape, with a border of ones separating a \quotes{large} bottom-right area of zeros.
	
	While we have characterized some relationships between incidence matrix properties, the critical exponent, and the maximization of the extinction area, our understanding remains incomplete, particularly for complex empirical systems.
	The specific exponents $\gamma_c$ and $\gamma_{EA}$ are likely influenced by how structural properties and correlations interact in shaping the system’s robustness. 
	Our flexible framework seems able to capture these relationships by weighting node importance in different ways, depending on the  system in analysis.  Moreover, the presence of a critical behavior raises the possibility that the model could capture complex non-local  structures in its proximity. 
	However, interpreting the final exponent values we identify is not trivial.

	The main limitation of our approach with respect to simpler algorithms is that there is an additional hyper-parameter $\gamma$ that have to be optimized depending on the task. This introduces a trade-off between computational cost and the precision at which the user wants to identify the ranking that, for instance, maximizes the extinction area. For small enough systems, directly setting $\gamma=1$ is typically a safe choice to maximize the extinction area,  but for minimizing the  nestendness temperature different exponents can be needed. For large ecosystems, the optimal $\gamma$ is often different from  -1 in both tasks, and thus the cost-precision trade-off has to be considered.

	Although we focused on ecological systems, it would be valuable to explore the applicability of our generalized map in other fields. 
	For instance, several results have been obtained using the fitness-complexity algorithm (i.e.,  the $\gamma=-1$ case) in economic systems \cite{tacchella2012new, cristelli2013measuring, cristelli2015heterogeneous, servedio2018new,tacchella2018dynamical}.
	In particular, there is an  interesting ongoing debate on the meaning and the evaluation of the  \quotes{complexity} of exported products and on the \quotes{fitness} of the exporter nations \cite{morrison2017economic, sciarra2020reconciling, balland2022reprint}. Different economic definitions would ideally  correspond to different values of $\gamma$ and can thus lead to different rankings of nations.

	Besides its practical applications, our non-linear map represents a relatively simple model with a rich phenomenology and  an emergent  "dynamical" phase transition that could be explored in more depth with statistical physics tools. For example, understanding the  deep connection between the condensation phase transition and the maximum of the extinction area,  shown by \FIG{fig:pht}c, is still an open issue.
	These two features seem closely  connected by the triangular shape of the incidence matrix. However the actual proof that the condensation is necessary for  the maximal extinction area (or the maximal that our mapping can provide) is still missing. 
	The possible connection between the condensation we observe and the well studied condensation transition in  network theory is also an open problem \cite{Bianconi2001}.
	Moreover, the recently discovered analogy between the fitness-complexity map and the Sinkhorn–Knopp algorithm~\cite{mazzilli2024equivalence} suggests a possible alternative direction to frame and understand the properties of our non-linear map.

	Our framework represents only one possible  generalization of ranking algorithms based on non-linear maps  as the fitness-complexity map \cite{tacchella2012new}.  A previous attempt along this line  \cite{mariani2015measuring} generalized the map in a non-symmetric manner and  can actually be easily included in our framework by re-defining the exponents for the $x$ and $y$ node types, i.e.,  $\gamma_x=-\beta$ and $\gamma_y=-1/\beta$. 
	However, our symmetric choice in  \EQ{eq:gen_pn_map}  leads, for example, to  larger extinction areas as analyzed in more detail in \SM{7}.
	The behavior of our map for $\gamma \rightarrow - \infty$ discussed in the method section is also reminiscent of the \quotes{Minimal Extremal Metric} introduced in ref.~\cite{wu2016mathematics}, where the score of the second node set is given by the minimal score of the first set. However, this extremal metric is obtained as the limit for large $\gamma$ of the different non-symmetric map discussed above~\cite{mariani2015measuring}.  In particular, differently from our case,  the score on one set of nodes  is still computed as in the fitness-complexity map.

	A more general extension of our map, which we plan to investigate, involves a two-parameter version where the non-linearity differs between the two node sets, i.e., $\gamma_x \neq \gamma_y$. In this case,  the  $\gamma_x,\gamma_y$ plane could be explored to select the best combination of  exponents given the task at hand. 
	This generalization  would allow the introduction of different  principles of node importance on the two sets with the consequent coupled rankings. The  statistical properties of this more general map in terms of convergence and phase transitions could present an even richer phenomenology.

	Finally, our approach can also be extended to  study node centrality on large multiplex networks  \cite{battiston2017,Bianconi2018}.
	Multiplex networks are playing an increasingly  important role in modern big data analyses. Many of these networks could be represented as multi-partite networks and could thus be approached with an extension of our map. 
	Moreover, multiplex networks implicitly  have a natural   bipartite scheme with the nodes on one side and the layers of the multiplex on the other side.  Weighted links connect a node with a layer with a weight  proportional to the number (and centrality) of neighboring nodes.  
	Following \cite{Bianconi2018},  we may rank the importance of nodes by increasing the centrality of nodes that receive links from highly influential layers,  and symmetrically enhance the influence of a layer if it contains  central nodes. 
	This idea naturally leads to an iterative map, but extended to include a weighted adjacency matrix.  
	An additional non-trivial extension of our approach could be developed for temporal networks \cite{holme2012},  adding yet another layer of dynamics to  the non-linear map.

	\section{Code availability}
	
	The node scores at the basis of our results can be computed with the software in the repository \url{https://github.com/amazzoli/xymap}.
	It contains also the code to generate the panels in Fig. \ref{fig:ext_area}d, \ref{fig:pht}a and \ref{fig:pack}.
	The code to generate all the other figures are available upon request.

	\onecolumn
	
	\bibliographystyle{unsrt}
	\bibliography{draft}
	
	\newpage

	\begin{center}
		\textbf{\large SUPPLEMENTARY MATERIAL}
	\end{center}

\setcounter{figure}{0}
\setcounter{section}{0}
\setcounter{table}{0}
\setcounter{equation}{0}

\renewcommand{\figurename}{Supplementary Figure}
\renewcommand{\thesection}{Supplementary Note \arabic{section}}
\renewcommand{\tablename}{Supplementary Table}

\section{Computational issues}

The $x$/$y$-scores are provided by the dynamical iteration of the map 
%{(1)}.
(1) of the main text.
In this Supplementary Note we discuss some technical points and approximations that are necessary to carry out a reliable numerical simulation of it.

\subsection{Dealing with the zeros}

A straightforward simulation of the $x$-$y$ map has the problem that some of the scores can have zero as fixed point. 
If $\gamma < 0$, those scores have to be elevated to a negative power, implying that some approximations must be introduced to deal with infinite values.
To understand the correct way to keep into account those zeros, let us go through the simple example of Figure \ref{fig:approx}.

\begin{figure}[h]
	\centering
	\includegraphics[width=0.9\textwidth]{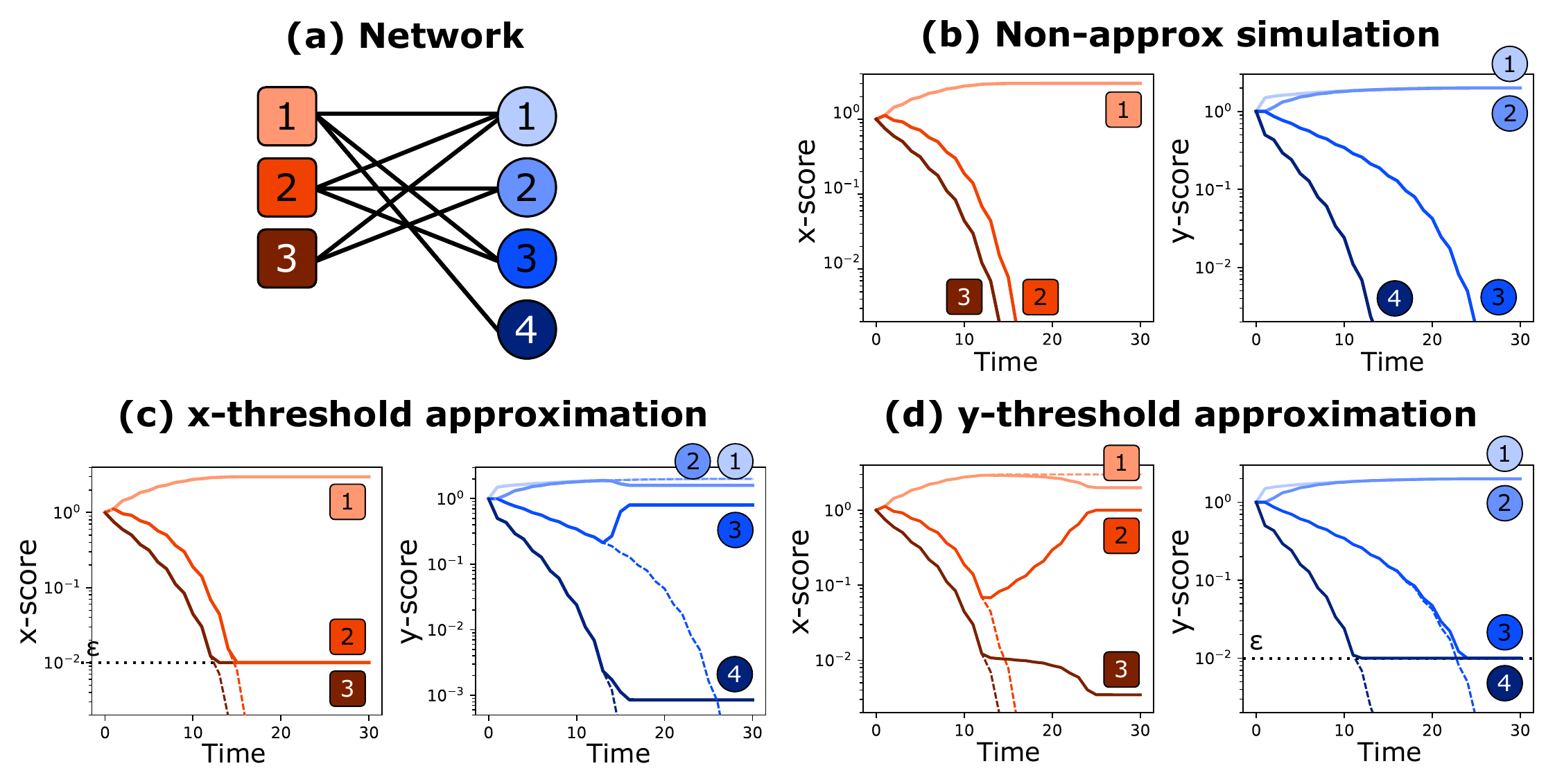}
	\caption{Trajectories of the $x$ and $y$ scores for the network in panel (a), evaluated with the algorithm at $\gamma=-1.2$.
		Panel (b) shows the non-approximated algorithm, panel (c) the map with a threshold $\epsilon$ imposed on the $y$ score, while panel (d) on the $x$ score.
		In panel (c) and (d) the dashed lines are to the non-approximated trajectories.}
	\label{fig:approx}
\end{figure}

The trajectories for a negative $\gamma$ without any approximation is shown in panel (b). 
Note that we can run the map without approximation only because the system is small and the simulation runs for a few time steps.
This shows one positive fixed point and two zeros for the $x$-scores, while two positive and two zeros for the other layer.
In general, the trajectories that go to zero will eventually reach  the numerical limit and they will be approximated with a zero, leading to a division by zero.
One possible solution is that, once a trajectory reaches a small value $\epsilon$, it stops decaying and it is set equal to $\epsilon$ for all the remaining time.
Panel (c) shows this procedure for a threshold $\epsilon$ applied only to the $x$-scores: $x_i \rightarrow \max \{ x_i, \epsilon\}$. The $y$-scores are computed without approximations.
Looking at the panel, it seems that $x$ scores follow the true trajectories (represented by dashed lines) until the threshold is reached.
In that case, the ranking can be computed putting the scores that reach the threshold first at lower positions in the ranking.
However, the $y$ scores show a weird behavior, and, in particular, the node number $3$ converges to a positive value, while the panel (b) says that it should go to zero.
In panel (d) the threshold is imposed to the $y$ scores, showing a reliable approximation for the $y$s, but not for the $x$s.
This picture does not change for the more complex case of Figure \ref{fig:approx_large}a, which shows the good approximation of $x$-thresholding for the $x$-scores, while an unreliable set of trajectories for the $y$ scores (with changes in the ranking with respect the true computation), and similarly for $y$-thresholding in panel \ref{fig:approx_large}b.

\begin{figure}[h]
	\centering
	\includegraphics[width=0.9\textwidth]{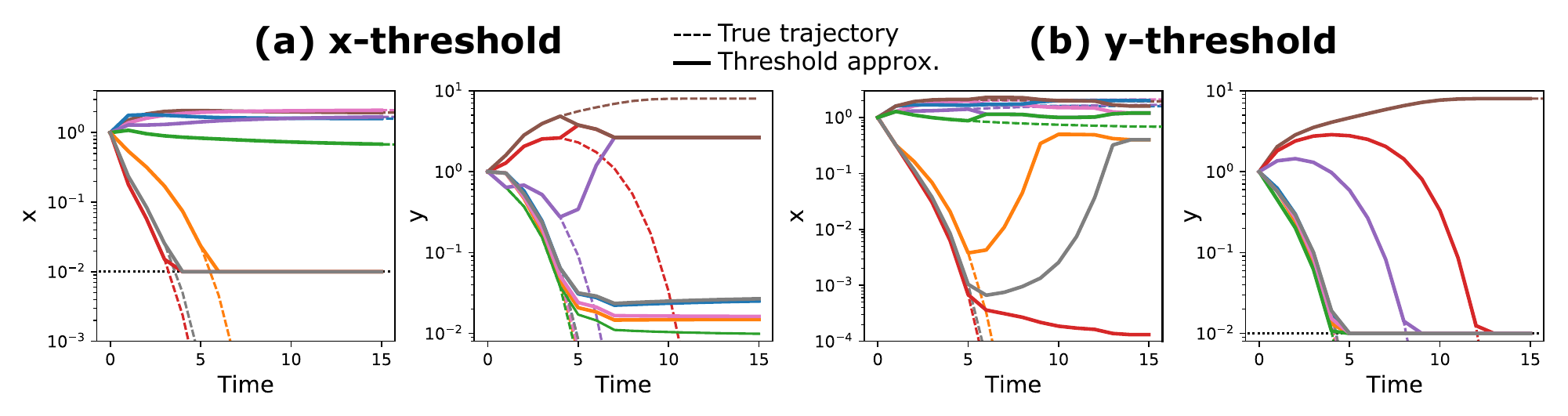}
	\caption{Threshold approximation for a $8 \times 8$ matrix. The $x$-scores for the $x$-thresholding and the $y$-scores for the $y$-thresholding overlap almost perfectly the non-approximated trajectories (dashed lines).}
	\label{fig:approx_large}
\end{figure}

\subsubsection{Wrong approximation of the $y$-score by the $x$-thresholding}

To better understand this behavior, let us first focus on why the $x$-threshold approximation does not work well for the $y$ score.
The same reasoning can be transposed to the $y$-threshold.
For example, we can focus on the $y$-trajectory of node $3$ in panel \ref{fig:approx}c.
This trajectory should go to zero, as panel (b) shows.
Indeed, in the regime where $x_1 \simeq 3 \gg x_2 \gg x_3$ (reached after few time steps), the score at the next step $y_3'$ as a function of the score two steps before $y_3$, reads:
\begin{equation*}
	y_3' = \frac{x_1^\gamma + x_2^\gamma}{\frac{1}{3}(3 x_1^\gamma + 3 x_2^\gamma + 2 x_3^\gamma)}\simeq 2 \left( \frac{x_2}{x_3} \right)^\gamma \simeq 2 \left( \frac{y_3}{2}\right)^{\gamma^2} \propto y_3^{\gamma^2}
\end{equation*}
where the third equality is obtained by using the expression of the $x$-scores as a function of the previous $y$-scores, and assuming $y_1 = y_2 = 2 \gg y_3 \gg y_4$.
Since $y_3$ is smaller than $1$ and $\gamma^2$ is a positive exponent, the trajectory will decay (more than exponentially) to zero.
This corresponds to the correct behavior, however when both $x_3$ and $x_2$ hit the threshold $\epsilon$ ($x_1 = 3 \gg x_2 = x_3 = \epsilon$), the value of $y_3'$ becomes constant and equal to $y_3 = 4/5$, drastically changing the fixed point value.
Also the two larger scores $y_1, y_2$ deviate from the true fixed point $2$ and converge to $8/5$.

In general, the thresholding procedure affects the smallest $x$-scores, but, for $\gamma < 0$, the smallest $x$-scores are also the most important contributions in computing the $y$-scores (the largest $y$s are the ones connected with the smallest $x$s).
In the next section it is shown that, quite surprisingly, even though these scores are wrong, the $x$-scores computed with the $x$-thresholding are instead well approximated.

\subsubsection{Good approximation of the $x$-score by the $x$-thresolding}

Let us write down the trajectories of the non-approximated $x$s in the regime $x_1 \simeq 3 \gg x_2 \gg x_3$:
\begin{equation}
	\begin{cases}
		x_1' \simeq 3
		\\
		x_2' \simeq 3 \left( \frac{x_2}{3} \right)^{\gamma^2}
		\\
		x_3' \simeq 3 \; 2^\gamma \left( \frac{x_3}{3} \right)^{\gamma^2}
	\end{cases}
	\label{eq:x_true}
\end{equation}
Where, the score $x_i'$ is expressed as a function of the score two steps before, $x_i$, using the map two consecutive times: the first to compute the $y$s as a function of the $x$s, and the second to get $x_i'$ as a function of the $y$s.
Considering Figure \ref{fig:approx}b, this behavior is followed until $x_3$ hits the threshold, let us say, at time $t_1$.
For $t > t_1$ this score is then imposed to be $x_3 = \epsilon$.
We also define $t_2$ as the time at which the trajectory $x_2$ becomes smaller than $\epsilon$ and it is set equal to $\epsilon$.
Between $t_1$ and $t_2$ the trajectories can be computed applying the map two times as before, but having $x_3 = \epsilon \ll 1$ and the condition $x_2 \gg x_3$ no longer satisfied.
The result is:
\begin{equation}
	\begin{cases}
		x_1' \simeq 3
		\\
		x_2' \simeq \frac{3}{3^{\gamma^2}} \left( 2(x_2^\gamma + \epsilon^\gamma)^\gamma + x_2^{\gamma^2} \right)
		\\
		x_3' = \epsilon
	\end{cases}
\end{equation}
Therefore, the largest score is not affected at all, while $x_2$ changes its trajectory.
However this change is negligible as long as it stays away from the threshold $x_2 \gg \epsilon$ (the equation above for the correct behavior can be easily recovered in this regime).
When $t > t_2$, and $x_2= x_3 = \epsilon$, one can see that the map provides again the right value for $x_1 \simeq 3$.
As a conclusion, the trajectories are well approximated as long as the scores are larger than the threshold. 
The fact that the approximation for the $x$s is good even though the $y$s used to compute them are wrong (as shown in the previous section) can be explained as follows. 
The threshold imposed to $x_3$ changes significantly the values of $y_1$ and $y_2$ at which it is connected, as we saw before.
However, the behavior of $x_2$ is dominated by $y_3$ ($y_3^\gamma$ is the largest term in its summation), while $x_1$ is driven by $y_4$.
The scores $y_1$, $y_2$ affect significantly only $x_3$, which has already reached $\epsilon$.
As a consequence, the error in $y_1$, $y_2$ does not \quotes{propagates} to the $x$-scores that are larger than the threshold.
Similarly, when $x_2$ hits the threshold, $y_3$ is now affected, but $x_1$ depends mostly on $y_4$.
This can seem a lucky case, but in fact this is due to the structure of the ordered adjacency matrix when there are zeros as fixed points discussed in Supplementary Note \ref{sec:cond_area}.
Referring to the Figure \ref{fig:adj_matrix_sketch} positions, the first set of nodes that hit the threshold is $\mathcal{N}_k$.
This strongly affects the values of the $y$-scores of $\mathcal{M}_0$ which are the only ones connected to them.
However, because of the matrix structure, all the other nodes not in $\mathcal{N}_k$ are connected to nodes with smaller $y$ (belonging to $\mathcal{M}_{j>0}$), and therefore the contribution of the wrong estimates is then negligible for the computation of the scores in $\mathcal{N}_{i<k}$.
This reasoning can then be applied iteratively for $\mathcal{N}_{k-1}$ hitting the threshold and modifying the $y$-scores in $\mathcal{M}_1$. As before, all the nodes in $\mathcal{N}_{i>k-1}$ have at least a connection with a node in $\mathcal{M}_{j>1}$ and the error does not propagate.

In conclusion, imposing a threshold to one of the two layers allows us to carry out a reliable simulation of the scores of that layer. 
The scores of the other layer have to be obtained with an independent simulation with the suitable thresholding procedure.

\subsection{Convergence criterion}

Another important point is to identify the stationary condition, at which the map iteration stops, and the scores are obtained.
The following simple procedure for $x$ is employed (and similarly for $y$): at each computational time step, the average differential of the fitness trajectories is computed, and reads as follows:
\begin{equation}
	\langle d_x \rangle_t = \frac{1}{n} \sum_{i=1}^n \mid x_i^{(t)} - x_i^{(t-2)} \mid
	\label{eq:conv}
\end{equation}
where $t$ is the map time-step.
Note that we are using the difference with the scores in $t$ and two steps before.
This is because computing the $x$-score every two steps results in a much faster simulation.
See the section \ref{sec:pseudocode} for a pseudocode of the algorithm.
The map stops if one of the two following conditions are satisfied: $\langle d_x \rangle_t < \theta$, where the threshold is chosen equal to $10^{-6}$, or if the number of iterations exceeds a limit value (about $2000$ iterations).
The latter condition is required in particular rare cases where the relaxation time becomes extremely large, such as if there are power law decaying trajectories, or when a stationary state does not exists.

\subsection{Computing the ranking}
\label{sec:ranking}

A third crucial point is the ranking computation, i.e. the ordering of the nodes based on the map score.
In a scenario where all the score trajectories converge to non-zero values, the ranking can be computed trivially, and it is just given by the sorted scores.
However, the presence of trajectories which decay to zero is very common. 
Clearly these nodes cannot be compared on the basis of their stationary value, which is zero (actually $\epsilon$) for all of them.
In such cases, we distinguish between the nodes with non-zero stationary fitness, and the nodes with trajectories decaying to zero and hitting the threshold $\epsilon$.
The nodes in the first group are put at the top positions of the ranking, and they are ordered according to their stationary score.
The zero trajectories are ranked according to the time at which they went below the boundary.
Specifically, the ones that crossed the threshold before are put at lower positions in the ranking.

\subsection{Pseudocode}
\label{sec:pseudocode}

The following pseudocode is based on the arguments discussed above.
An implementation in python of the algorithm can be found at the following repository: \url{https://github.com/amazzoli/xymap.git}.
Note that instead of using $x$ and $y$, we refer to $a$ as the score that we want to reliably compute, and $b$ the opposite score (that can be affected by the thresholding approximation).
\\

\textbf{Input:} $A$, adjacency matrix of the bipartite network.

\textbf{Input:} $\gamma$, exponent of the map.

\textbf{Input:} $a \in \{ x, y \}$, which scores to compute (\textit{axis}). 

\textbf{Define:} $b$, opposite score of $a$, $b=y$ if $a=x$, or $b=x$ if $a=y$.

\textbf{Input:} $\tau_{max}$, maximum number of allowed iterations.

\textbf{Input:} $\epsilon$, zero threshold for approximating the zeros of the $a$-score.

\textbf{Input:} $\delta$, average $a$-score differential below which the algorithm converges.
\\

\textbf{Initialize:} $\mathbf{a}^{(0)}$, $\mathbf{b}^{(0)}$, initial conditions of the scores as vectors of ones.

\textbf{Initialize:} $\mathbf{r}_a$, the ranking of the score $a$ as an empty vector.

\textbf{Initialize:} $\mathcal{Z}_a$, the set of nodes of the layer $a$ which have reached the threshold $\epsilon$ as an empty set.
\\

\textbf{Loop:} for $\tau = 1, \ldots, \tau_{max}$:

\hspace{0.5cm} $|\;$ Compute the opposite score: $\mathbf{b}^{(2\tau-1)} = A^T (\mathbf{a}^{(2\tau-2)})^\gamma$

\hspace{0.5cm} $|\;$ Compute the main score: $\mathbf{a}^{(2\tau)} = A (\mathbf{b}^{(2\tau-2)})^\gamma$

\hspace{0.5cm} $|\;$ Check which nodes have crossed the threshold in this step: $\mathcal{Z}_a^{\text{new}} = \{ i \; | \; \mathbf{a}^{(2\tau)}_i \le \epsilon \text{ and } i \not\in \mathcal{Z}_a \}$.

\hspace{0.5cm} $|\;$ Update the ranking with the new nodes below threshold: $\mathbf{r}_a \leftarrow ( argsort[ \mathbf{a}^{(2\tau)}_{i \in \mathcal{Z}_a^{\text{new}}}], \mathbf{r}_a )$ \footnote{\textit{$argosort[\mathbf{v}]$} returns the indexes $\{ i_1, \ldots, i_k \}$ of the vector $\mathbf{v}$ in the order such that $v_{i_1} \ge v_{i_2} \ge \ldots \ge v_{i_k}$.}

\hspace{0.5cm} $|\;$ Update the set of nodes below threshold $\mathcal{Z}_a \leftarrow \mathcal{Z}_a \; \bigcup \; \mathcal{Z}_a^{\text{new}}$

\hspace{0.5cm} $|\;$ Impose the threshold approximation: $a_i^{(2\tau)} = \epsilon$ if $i \in \mathcal{Z}_a$.

\hspace{0.5cm} $|\;$ Compute the quantity in equation (\ref{eq:conv}): $\langle d_a\rangle$

\hspace{0.5cm} $|\;$ Check the convergence: \textbf{If}  $\langle d_a\rangle < \delta$:

\hspace{0.5cm} $|\;$\hspace{0.5cm} $|\;$ \textbf{Break} the loop

\textbf{Loop end.}

Compute the ranking of the nodes with positive fixed point: $\mathbf{r}_a \leftarrow ( argsort[ \mathbf{a}^{(2\tau)}_{i \not\in \mathcal{Z}_a}], \mathbf{r}_a )$
\\

We refer to $t$ as the time of the map written in equation (1) of the main text.
$\tau$ is instead the time step of the loop which corresponds to $\tau = \text{int}(t/2)$.
Note that the algorithm computes the trajectories every two steps: $(\mathbf{a}^{2}, \mathbf{a}^{4}, \ldots)$ and $(\mathbf{b}^{1}, \mathbf{b}^{3}, \ldots)$.

\section{Derivation of the singular vector centrality in the case of $\gamma = 1$}

Here we prove that the $\mathbf{x}$/$\mathbf{y}$ scores converge to the left/right singular vectors of the incidence matrix corresponding to the leading singular value.
Let us consider the map written in matrix form:
\begin{equation*}
	\left\lbrace
	\begin{aligned}
		& \mathbf{x}^t = \frac{\tilde{\mathbf{x}}^{t}}{\langle \tilde{\mathbf{x}}^{t} \rangle}, \; \text{ where } \; \tilde{\mathbf{x}}^{t} = A  \mathbf{y}^{t-1} 
		\\
		& \mathbf{y}^{t} = \frac{\tilde{\mathbf{y}}^{t}}{\langle \tilde{\mathbf{y}}^{t} \rangle}, \; \text{ where } \; \tilde{\mathbf{y}}^{t} = A^T \mathbf{x}^{t-1}
	\end{aligned}
	\right. .
\end{equation*}
We now consider the proportionality between the $x$-score and the left singular vector (the derivation is similar for the $y$-score).
Writing down recursively the map one obtains:
\begin{equation*}
	\begin{aligned}
		\mathbf{x}^t = \frac{A \mathbf{y}^{t-1}}{\langle A \mathbf{y}^{t-1} \rangle} = \frac{A A^T \mathbf{x}^{t-2}}{\langle A A^T \mathbf{x}^{t-2} \rangle} & \; = \; t \text{ even} = \frac{(A A^T)^{t/2} \; \mathbf{x}^{0}}{\langle (A A^T)^{t/2} \; \mathbf{x}^{0} \rangle}
		\\
		& \; = \; t \text{ odd} = \frac{(A A^T)^{(t-1)/2} \; A \mathbf{y}^{0}}{\langle (A A^T)^{(t-1)/2} \; A \mathbf{y}^{0} \rangle}
	\end{aligned} , 
\end{equation*}
where $\mathbf{x}^0$ and $\mathbf{y}^0$ are the initial conditions of the map which we choose to be vectors of ones.
The $n \times m$ matrix $A$ can be then rewritten using the singular value decomposition: $A = U D V^T$, where $U = (\mathbf{u}^1, \ldots, \mathbf{u}^n)$ and $V = (\mathbf{v}^1, \ldots, \mathbf{v}^m)$ are unitary matrices whose columns are, respectively, the left and right singular vectors of $A$.
The matrix $D$ is a rectangular diagonal matrix: $D_{ij} = \sigma_i \delta_{ij}$.
Using the singular value decomposition in the equation above, one obtains:
\begin{equation*}
	\begin{aligned}
		\mathbf{x}^t & = \; t \text{ even} = \frac{U (D D^T)^{t/2} U^T \mathbf{x}^{0}}{\langle U (D D^T)^{t/2} U^T  \mathbf{x}^{0} \rangle}
		\\
		& = \; t \text{ odd} = \frac{U (D D^T)^{(t-1)/2} D V^T \mathbf{y}^{0}}{\langle U (D D^T)^{(t-1)/2} D V^T \mathbf{y}^{0} \rangle} . 
	\end{aligned}
\end{equation*}
Then one can consider the element $i$ of the following vector:
\begin{equation*}
	\left( U (D D^T)^{t/2} U^T \mathbf{x}^{0} \right)_i = \sum_{j,k,l,p} u_i^j \sigma_j^{t/2} \delta_{kl} \sigma_k^{t/2} \delta_{jk} u_p^l = \sum_{j} u_i^j \sigma_j^t \sum_l u_l^j , 
\end{equation*}
We define $\sigma_1$ the leading singular value, which the Perron-Frobenius theorem guarantees to be unique, positive and that the associated left/right singular vectors to be positive.
Therefore, by taking the limit of $t\rightarrow \infty$ one gets:
\begin{equation*}
	\begin{aligned}
		\left( U (D D^T)^{t/2} U^T \mathbf{x}^{0} \right)_i  & \rightarrow  u_i^1 \sigma_1^t \sum_l u_l^1 
		\\
		\left( U (D D^T)^{t/2} D V^T \mathbf{y}^{0} \right)_i  & \rightarrow u_i^1 \sigma_1^t \sum_l v_l^1 
	\end{aligned}
\end{equation*}
By putting everything together, one can express the $i$ component of the $x$-score in the large $t$ limit:
\begin{equation}
	x_i^{t} \rightarrow
	\begin{aligned} 
		&\; t \text{ even } \rightarrow \frac{u_i^1 \sigma_1^t \sum_l u_l^1 }{\frac{1}{n}\sum_i u_i^1 \sigma_1^t \sum_l u_l^1 } 
		\\
		&\; t \text{ odd } \rightarrow \frac{u_i^1 \sigma_1^t \sum_l v_l^1 }{\frac{1}{n}\sum_i u_i^1 \sigma_1^t \sum_l v_l^1 }
	\end{aligned}
	\; = u_i^1 \frac{n}{\sum_i u_i^1}
	\label{eq:x_sing}
\end{equation}

\section{Genetic algorithm}

To check the efficiency of the $x$-$y$ map in finding the best extinction area, we compare its results with a genetic algorithm over a set of random binary square matrices with growing size.
Let us define the matrix size $s$ as the number of rows and columns.
We have considered the two ensembles of random matrices.
Uniform random matrices in which is element has a given probability $p$ to be one, and a \quotes{heterogeneous} set, defined by the condition that an element $m_{ij}$ is equal to $1$ with probability $\frac{2 i j}{s(s+1)}$, where $i = 1, \ldots s$ and $j = 1, \ldots s$, and zero otherwise.
In this way the average of the row $i$ degree is $d_i = \sum_j p_{ij} = i$, and similarly $d_j = j$.
Figure \ref{fig:gen_alg} shows the computational time and the best extinction area found  as a function of the matrix {row and column number}, for the two algorithms and the standard fitness-complexity map.

The genetic algorithm starts with a population composed of $N$ row rankings generated at random (i.e. each one is a random permutation of the row indexes).
At each time step two rankings are drawn from the population and the two associated extinction areas are computed.
The row ordering with the lesser extinction area is substituted by a copy of the other "fitter" ranking, and this new copy can mutate with probability $\mu$, where a mutation means that two row indexes exchange their position in the ranking.
The algorithm stops if there is no improvement of the extinction area for $t_{stop}$ iterations.

It is also necessary to define a procedure to find the exponent of the $x$-$y$ map which maximizes the extinction area.
We employ two different methods that are compared.
The first one is the Nelder-Mead simplex algorithm \cite{nelder1965simplex} which is implemented in the \textit{scipy.optimize.fmin} method of python.
The second one is based on the bisection algorithm.
It is faster but assumes that the extinction area as a function of the map exponent is concave, which seems to be approximately true in a lot of empirical cases.
The method is the following: at a generic step $t$, the extinction area is computed for an exponent $\gamma_t$ and a second exponent $\gamma_{t+1} = \gamma_t + \Delta_t$.
If the extinction area at $\gamma_{t+1}$ is greater than the area of the first exponent, this means that I am approaching the maximum, and, at the next time step, $\gamma_{t+2} = \gamma_{t+1} + \Delta_{t+1}$, with $\Delta_{t+1} = \Delta_{t}$.
Otherwise, if the area at $\gamma_{t+1}$ is smaller, the increment becomes negative and divided by a factor $2$: $\Delta_{t+1} = - \frac{\Delta_{t+1}}{2}$.
The method stops if the exponent increment becomes smaller than a certain threshold $\hat{\Delta}$.

\begin{figure}[h]
	\centering
	\includegraphics[width=\textwidth]{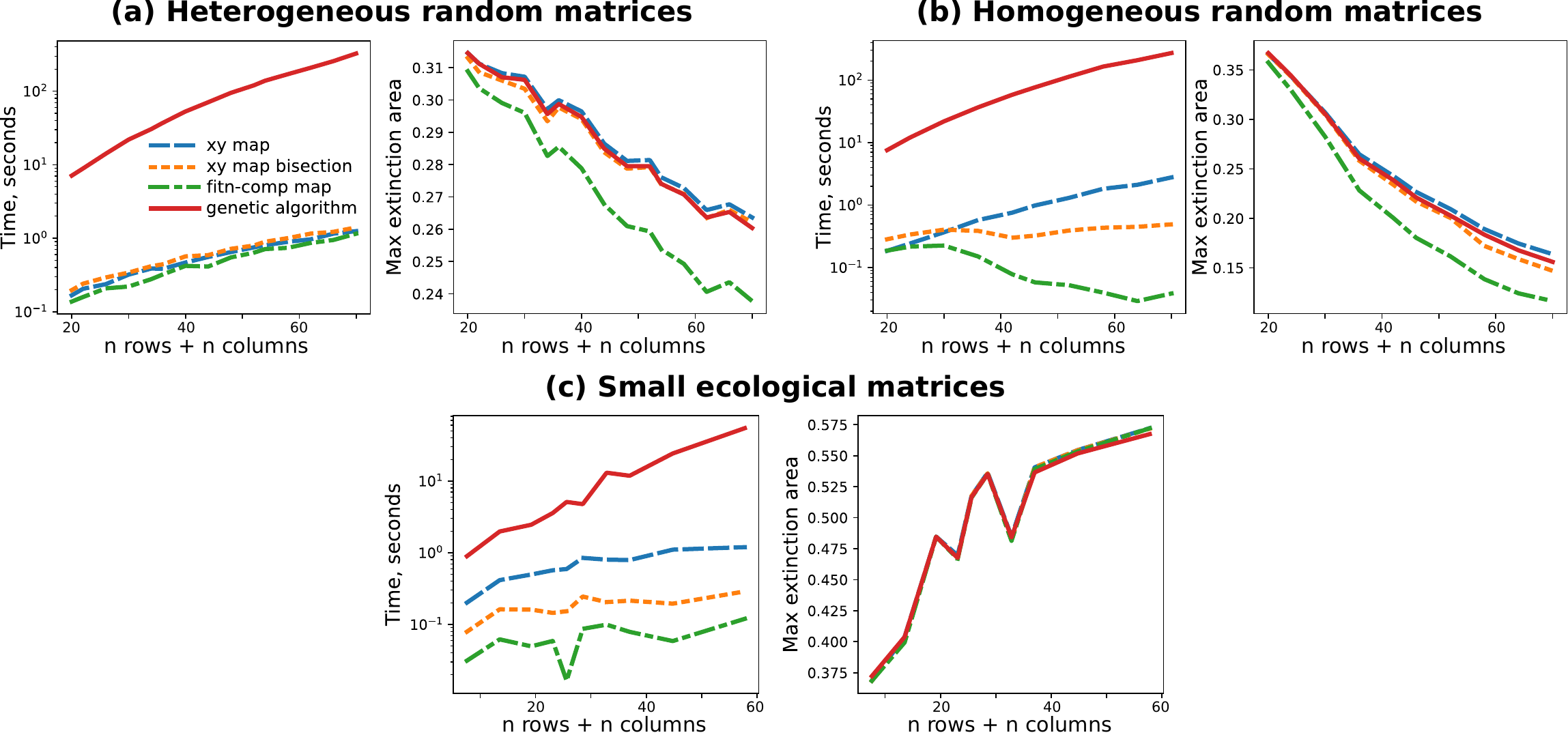}
	\caption{Comparison of a genetic algorithm, the $x$-$y$ map (using two procedures for finding the optimum), and the fitness-complexity map for the performance in computing the extinction area of a bipartite system.
		Two ensembles of random matrices have been considered whose precise definition is in the main text: uniform matrices with $p=1/3$, panel (a), and heterogeneous matrices panel (b).
		The left plot of each panel shows the computational time (in seconds) as a function of the matrix dimension, while on the right there is the best extinction area found by each algorithm.
		The two quantities are the average over an ensemble of $200$ random matrices for each size.
		The definition of the algorithms and procedure of the matrix generation are explained in the main text.
		The parameters of the genetic algorithm are: $N = 500$, $\mu = 0.05$, and $t_{stop} = 10^4$, while for the bisection of the $x$-$y$ map: $\gamma_0 = 0.8$, $\Delta_0 = 0.2$, $\hat{\Delta} = 0.01$.
		{Panel (c): same computation for the small ecological network of the Web of Life database, where we selected the matrices having size (product of number of rows and number of columns) smaller than $600$.
			In total we consider $185$ binary matrices.
			The values of time and extinction area are an average over a given bin of number of rows plus number of columns, typically containing around 20 points.}
	}
	\label{fig:gen_alg}
\end{figure}

Figure \ref{fig:gen_alg} shows that the $x$-$y$ map with the simplex algorithm finds, on average, a better extinction area than all the other algorithms. The fact that it is even larger than the genetic algorithm should be due to the very large space of the ranking that cannot be reliably explored in reasonable computational times.
The bisection method applied to the $x$-$y$ map can be faster than the simplex method (panel (b)), but can be trapped in local maxima and performs a bit poorer, suggesting that the extinction area as a function of the exponent $\gamma$ is not always perfectly concave.
The fitness-complexity map performs significantly worse than the other algorithms.
It is curious that the computational time decreases with the size in homogeneous matrices.
This happens because the transition point, which is close to one for small matrices, increases with the size, and, as a consequence, the computational time to run the map farther from the transition decreases.

\section{Computing the phase transition exponent}

From Figure 4 of the main manuscript and Figure \ref{fig:pht_all}, it is clear that the phase transition of the fraction of positive scores is not always a sudden jump from $1$ to a number $O(1/n)$.
Therefore, to numerically compute the transition exponent one needs a general procedure that gives reliable estimates in all the possible scenarios.
We employed the following algorithm:
\begin{itemize}
	\item Find the largest exponent for which the number of convergent trajectories is $O(1/n)$, $\gamma_{\text{left}}$.
	To this end, we start simulating the map at a very low exponent, compute the fraction of positive scores, $f_{\text{small}}$, and iteratively increase the exponent using a bisection method to find the point in which a number of positive scores larger than $f_{\text{small}}$ converges.
	\item Find the smallest exponent for which the number of convergent trajectories is $1$, $\gamma_{\text{right}}$. A similar procedure of the step above is employed.
	\item Compute the fraction of positive scores, $\{f_1, \ldots, f_k \}$ for a set of evenly spaced exponents between $\gamma_{\text{left}}$ and $\gamma_{\text{right}}$: $\{\gamma_1, \ldots, \gamma_k \}$.
	\item The critical exponent is the average of the computed gammas weighted by the size of the jumps between them: 
	$$
	\gamma_c = \frac{\sum_{i=1}^{k-1} \frac{(\gamma_{i+1}+\gamma_i)}{2} |f_{i+1}-f_{i}|}{\sum_{i=1}^{k-1} |f_{i+1}-f_{i}|} 
	$$
	
\end{itemize}

\begin{figure}
	\centering
	\includegraphics[width=\textwidth]{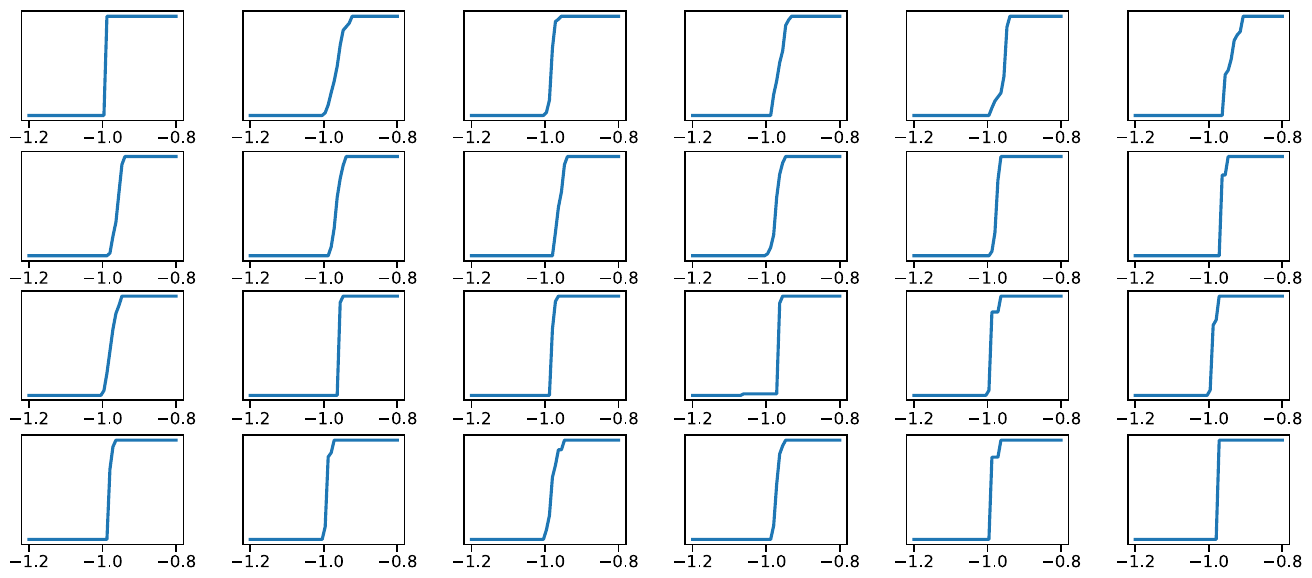}
	\caption{Fraction of convergent trajectories as a function of the map exponent in different ecological mutualistic network dataset. 
		Specifically the $24$ largest systems.
		All the datasets display a phase transition in the number of positive stationary trajectories, and the critical exponent varies.}
	\label{fig:pht_all}
\end{figure}

\section{A stable solution in the condensation phase leads to the bottom-right area of zeros}
\label{sec:cond_area}

For a large enough negative value of $\gamma$, the systems lies  in the condensed phase and the distribution of scores is characterized by a few scores of order $1$ while all the remaining scores
decay to zero exponentially or more than exponentially, see Figure 4 of the main text.
This distribution of scores can be parameterized as follows:
\begin{equation}
	\mathbf{x} = (\boldsymbol{\epsilon}_{0}, \boldsymbol{\epsilon}_{1}, \ldots, \boldsymbol{\epsilon}_{k})
	\label{eq:cond_score}
\end{equation}
where $\boldsymbol{\epsilon}_{k}$ denotes a set of scores of order $\epsilon_k$ and the few scores of order $1$ are in the first block $x_i \in \boldsymbol{\epsilon}_{0}$.
Let us call $\mathcal{N}_p = \{ i | x_i \in  \boldsymbol{\epsilon}_{p} \}$ the set of indexes of the scores that belong to the block $\boldsymbol{\epsilon}_{p}$, and its size $n_p = |\mathcal{N}_p|$.
Moreover let us assume that each block $\boldsymbol{\epsilon}_{p}$ is characterized by an order of magnitude $\epsilon_{p}^{(t)}$ which is much larger than the order of magnitude of the next block, namely, for a sufficiently larger $t$:
\begin{equation}
	\frac{x^{(t)}_{i \in \mathcal{N}_{p+1}}}{x^{(t)}_{j \in \mathcal{N}_p}} \sim \frac{\epsilon_{p+1}^{(t)}}{\epsilon_{p}^{(t)}} \ll 1
	\label{eq:order_separation}
\end{equation}
For the moment, $k$ can be whatever integer larger than $0$, since at least two blocks exist in the condensed phase: the positive scores and the ones going to zero.
It will be shown later that, for consistency, the number of scores in each block is of order $1$, implying $k$ to be of the same order of the matrix size.

The only requirement we must impose on eq. \ref{eq:cond_score} if we want to have a consistent solution of the map in the condensed phase is that the pattern of scores is stable under the mapping, i.e. that it reproduces itself after a complete iteration of the mapping.
More precisely, a stable solution would require that the scores belonging to the first block are stationary, $x_i^{(t+2)} = x_i^{(t)}$ $ \forall i \in \mathcal{N}_0 $, while all the others decrease,  $x_i^{(t+2)} < x_i^{(t)}$ $\forall i \notin \mathcal{N}_0 $, and the property \ref{eq:order_separation} is conserved.
What we are going to study now is the $y$ obtained from such an assumption, which conditions are implied by imposing the stability, and which shape the ordered adjacency matrix has in this regime.

The first step is to compute the scores in the opposite layer $\mathbf{y}^{(t+1)}$ given $\mathbf{x}^{(t)}$ having the shape (\ref{eq:cond_score}).
Let us call the set of nodes in the second layer connected with the smallest $x$-scores (i.e. the block $\boldsymbol{\epsilon}_{k}$) as $\mathcal{M}_0 = \{ j | \exists i \in \mathcal{N}_k | A_{ij} = 1 \}$.
We also call the $y$-scores of the nodes in this set as $\boldsymbol{\eta}_{0} = (y_1, \ldots, y_{m_0})$.
Thanks to the separation of scales (\ref{eq:order_separation}), and the fact that $\gamma < 0$, those scores can be approximated as:
\begin{equation}
	\tilde{y}_{j \in \mathcal{M}_0}^{(t+1)} \simeq \sum_{i \in \mathcal{N}_k} A_{ij} \left(x_i^{(t)}\right)^\gamma \simeq \left(\epsilon_k^{(t)}\right)^\gamma d_j^k
\end{equation}
where, for a more compact notation we introduced $d_j^k = \sum_{i \in \mathcal{N}_k} A_{ij}$, which is the number of links from $j$ to a node in the block $k$ of the first layer.
%The nodes in this set are clearly the largest ones, since the score of every node is dominated by the smallest $x$-score at which they are connected, and $\epsilon^k$ is the order of the smallest one.
One can then iterate the procedure and define $\mathcal{M}_1$ as the nodes connected with  $\boldsymbol{\epsilon}_{k-1}$ and not with $\boldsymbol{\epsilon}_{k}$, leading to a score of order $(\epsilon_{k-1}^{(t)})^\gamma$.
In general $\mathcal{M}_q = \{ j \notin \mathcal{M}_{q-1}| \exists i \in \mathcal{N}_{k-q} | A_{ij} = 1 \}$, for $q=0, \ldots, k$, defines the nodes whose lowest $x$ neighbour is of order $\epsilon_{k-q}^{(t)}$.
Note that, at this point, there is no guarantee that $\mathcal{M}_q$, for $q>1$, contains some nodes, however the analysis can be performed without problems allowing for empty sets.
It will be found later that the stability of (\ref{eq:cond_score}) requires that all the sets $\mathcal{M}_q$ contain at least one element.
The score of the nodes belonging to $\mathcal{M}_q$ (after the normalization step) read:
\begin{equation}
	y_{j \in \mathcal{M}_q}^{(t+1)} = \frac{\tilde{y}_{j \in \mathcal{M}_q}^{(t+1)} }{\langle \tilde{\mathbf{y}}^{(t+1)} \rangle} \simeq \frac{M d^{k-q}_j}{\sum_{j\in \mathcal{M}_0} d_j^k} \left( \frac{\epsilon_{k-q}^{(t)}}{\epsilon_k^{(t)}}\right)^\gamma
	\label{eq:cond_y_score}
\end{equation}
Therefore, one can see that also the ordered $y$-scores form a block structure $\mathbf{y} = (\boldsymbol{\eta}_{0}, \boldsymbol{\eta}_{1}, \ldots, \boldsymbol{\eta}_{k})$, where scores of order one are only contained in $\boldsymbol{\eta}_{0}$, and the other blocks are of order $\eta_q \sim M (\epsilon_{k-q}/\epsilon_k)^\gamma$.
Moreover, a counterpart of relation (\ref{eq:order_separation}) is satisfied: $\eta_{q+1}/\eta_q \ll 1$.

Now one can propagate the $y$-scores to the subsequent step of the $x$-score.
Let us define $\mathcal{N}_0'$ as the set of the first-layer nodes connected with the nodes in $\mathcal{M}_k$, the ones with the lowest $y$-score, of order $\eta_k^{(t+1)} \sim (\epsilon_k^{(t)})^{-\gamma}$.
This leads to:
\begin{equation}
	\tilde{x}_{i \in \mathcal{N}_0'}^{(t+2)} \simeq \sum_{j \in \mathcal{M}_k} A_{ij} \left(y_i^{(t+1)}\right)^\gamma \simeq \left(\eta_k^{(t+1)}\right)^\gamma d_i^k = \left(\epsilon_k^{(t)}\right)^{-\gamma^2} d_i^k
\end{equation}
Since those $x$-scores are much larger than all the other scores, the average $\langle \tilde{\mathbf{x}}^{(t+2)} \rangle$ is dominated by them, and it is of the same order.
This implies that the nodes in $\mathcal{N}_0'$ are of order $1$, and, by imposing the stationary condition of the solution, they must coincide with the order-$1$ scores at time $t$.
Therefore $\mathcal{N}_0 = \mathcal{N}_0' = \{ i | \exists j\in \mathcal{M}_k | A_{ij}=1 \}$.
As a consequence, the ordered adjacency matrix will show a block connecting $\mathcal{N}_0 $ and $\mathcal{M}_k $ at the top-right corner, see Figure \ref{fig:adj_matrix_sketch}, and all the other elements below that block are zero (no other nodes of the first layer are connected to nodes in $\mathcal{M}_k$).
In a similar way one can define $\mathcal{N}_1'$ as the nodes connected with $\mathcal{M}_{k-1}$ and not with $\mathcal{M}_{k}$.
These nodes are of the second leading order, and, to be consistent with a stationary solution of the kind (\ref{eq:cond_score}), they must coincide with the nodes in $\mathcal{N}_1$.
Iterating this procedure one obtains that $\mathcal{N}_p = \mathcal{N}_p' = \{ i \notin \mathcal{N}_{p-1} | \exists j\in \mathcal{M}_{k-p} | A_{ij}=1 \}$.
Note that, to be consistent with the stationary condition, all the $\mathcal{M}_q$, $q=1, \ldots, k$ must be non empty, otherwise, if $|\mathcal{M}_q| = 0$, the nodes in $\mathcal{N}_{k-q}'$ are of the same order of the nodes in another block, going against the separation of scales (\ref{eq:order_separation}).
The consequence of this structure, where no scores in the block $\mathcal{N}_p$ are connected to scores in the blocks $\mathcal{M}_q$, $q > k-p$, implies the triangular shape of the ordered adjacency matrix shown in Figure \ref{fig:adj_matrix_sketch}.
The area of zeros below the diagonal of blocks $\mathcal{N}_p$-$\mathcal{M}_{k-p}$ is consistent with the behavior observed in {Figure 5 of the main text}.
The shape of the scores belonging to the block $p$ reads:
\begin{equation}
	x_{i \in \mathcal{N}_p}^{(t+2)} \simeq N \frac{B_i^p}{\sum_{i' \in \mathcal{N}_0} B_{i'}^0} \left(\frac{\epsilon_p^{(t)}}{\epsilon_0}\right)^{\gamma^2}
	\label{eq:cond_decay}
\end{equation}
where, for simplicity of notation we introduced $B_i^p = \sum_{j \in \mathcal{M}_{k-p}} A_{ij} \left( d_j^p \right)^\gamma$.

\begin{figure}[!h]
	\centering
	\includegraphics[width=0.35\textwidth]{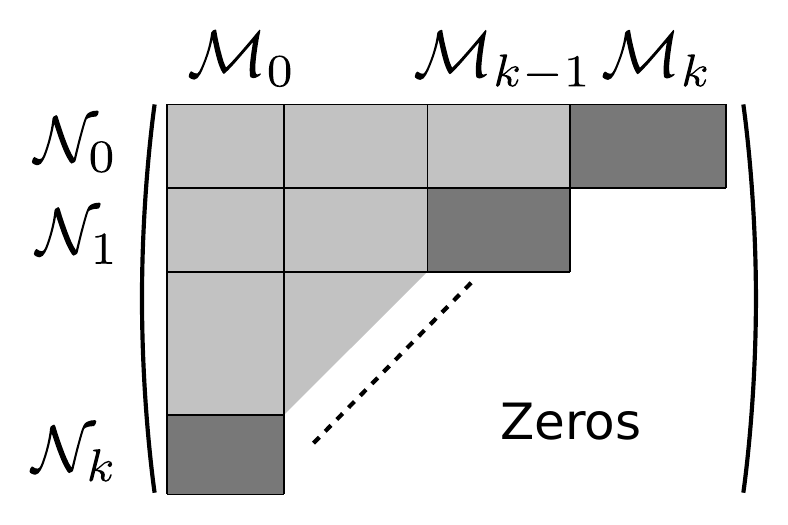}
	\caption{Adjacency matrix ordered according to the scores in the condensation regime.}
	\label{fig:adj_matrix_sketch}
\end{figure}

If one assumes that the $B_i^p$ are approximately of the same order of magnitude for each $p$ and $i$, the scores are: $x_{i \in \mathcal{N}_p} \sim N (\epsilon_p^{(t)}/ \epsilon_0)^{\gamma\cdot\gamma} /n_0$.
Under this approximation, the condition that each block $\boldsymbol{\epsilon}_p$, with $p>0$, decreases at each step, can be written as follows:
\begin{equation}
	\left( \frac{\epsilon_p^{(t)} n_0}{N}\right)^{\gamma^2-1} < 1
	\label{eq:mean_field}
\end{equation}
Since a decreasing $\epsilon_p^{(t)}$ becomes eventually less than $N$, the only possibility is to choose $\gamma < -1$, providing a qualitative upper bound for the condensed phase, equation (4) of the main text.

A further consequence of relation (\ref{eq:cond_decay}) is that each score in a block must have the same value, $x_i = x_j = \epsilon^p$, $\forall x_i, x_j \in \boldsymbol{\epsilon}^{p}$.
Indeed a small difference between two scores in the same block leads to a separation of scales after a sufficiently large number of steps.
If, for example, $x_i^{(t)} = \delta + x_{i+1}^{(t)}$, with $\delta > 0$:
\begin{equation*}
	\frac{x_{i+1}^{(t+2k)}}{x_i^{(t+2k)}} \propto \left( 1 + \frac{\delta}{x_{i+1}^{(t)}}\right)^{-\gamma^{2k}} \ll 1
\end{equation*}
Implying that two distinct blocks can be created and a new solution (\ref{eq:cond_score}) can be written.
This consideration has two important consequences.
First, this justifies \textit{a posteriori} the separation of the scales (\ref{eq:order_separation}) assumed at the beginning.
Second, it implies that all the nodes in the same block must have the same value, and this happens only if they are connected exactly to the same nodes, which happens very rarely in empirical networks.
In turns, we expect that the sizes of the blocks are $1$ or a few units.
The sub-matrices $\mathcal{N}_p$-$\mathcal{M}_{k-p}$ that separate the area of zeros form the rest of the matrix are, therefore, very small, and this can explain the \quotes{border of ones} that one typically sees in the ordered matrix in the condensation phase.

\section{Adjacency matrix shape and extinction area}
\label{sec:ext_area_adj_mat}

It can be proved that the number of consecutive zeros from the bottom of the matrix (Figure \ref{fig:pack_ext_area}a, green area) is proportional to extinction area computed removing rows (\ref{fig:pack_ext_area}b).
Also the transposed relation is true: the area of contiguous zeros from the matrix right side is proportional to the extinction area calculated by removing columns.

\begin{figure}[!h]
	\centering
	\includegraphics[width=0.45\textwidth]{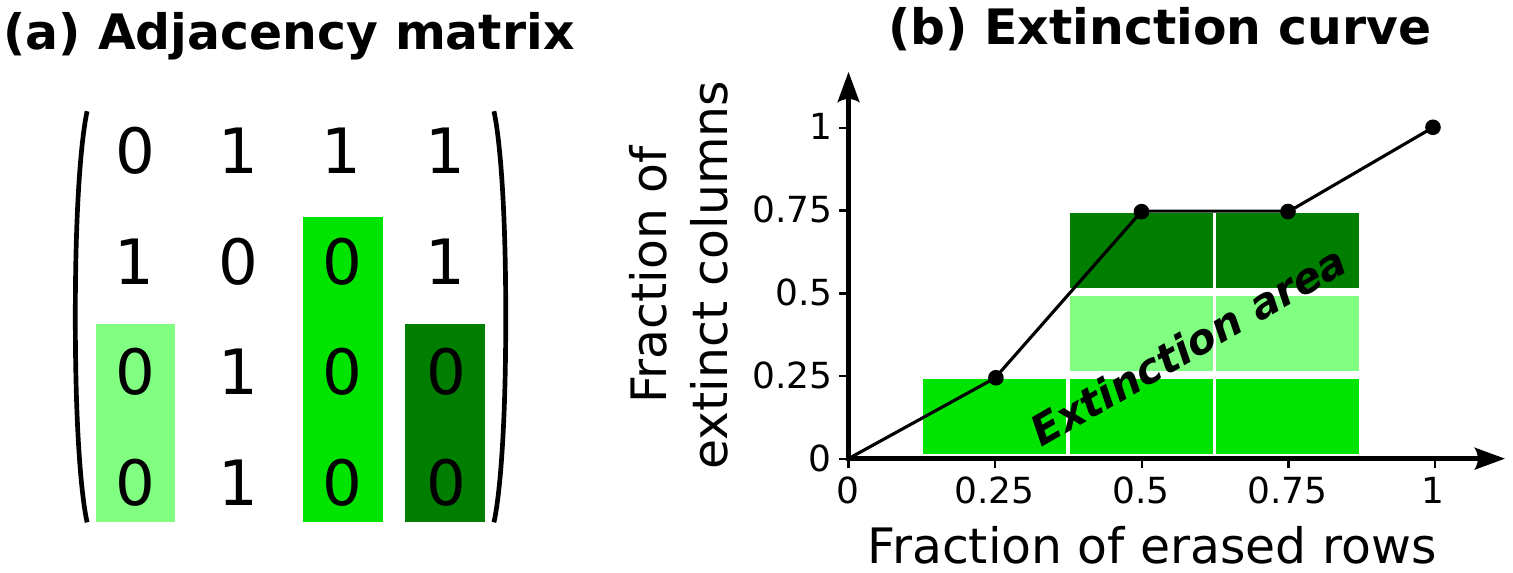}
	\caption{The figure shows the equivalence between the area of zeros from the matrix bottom, panel \textit{(a)} and the extinction area computed removing rows, panel \textit{(b)}.
		If the first row is removed, the third column gets extinct, and therefore the extinction curve increases by one green block at \quotes{fraction of erased rows}$ = 0.25$.
		Note that this contribution to the curve is represented by a box with the same shade of green of the extinct column.
		The removal of the second row leads to the extinction of the first and the fourth columns, therefore the height of the extinction curve is now at three blocks.
		Iterating the procedure one can see the equivalence between the area of zeros and the extinction area.}
	\label{fig:pack_ext_area}
\end{figure}

Let us start to formalize the area of zeros from the matrix bottom.
For each column $j$, the number of consecutive zeros from the bottom is $z_j$. It is also useful to define the following row index: $l_j = \max_i (A_{ij} = 1)$, which is the largest $i$ associated to a $1$ in the column $j$.
It is easy to see that $z_j = N - l_j$, where $N$ is the number of rows.
For example, in Figure \ref{fig:pack_ext_area}a, $l_1 = 2$, $l_2 = 4$, and so on.
The area of zeros is:
\begin{equation}
	Z = \sum_{j=1}^{M} z_j = \sum_{j=1}^{M} (N - l_j)
	\label{eq:A}
\end{equation}
where the sum is over all the $M$ column.
%Note that we have used the superscript $(c)$ to distinguish between the area from the matrix bottom in columns, and the area for the matrix right, $A^{(r)}$, i.e. the number of consecutive zeros form the right side of each row.
We now assume that the rows of the adjacency matrix are ordered according to the ranking at which we want to compute the extinction area.
In order to compute this quantity, the first node/row in the ranking is removed.
The node in the opposite layer which remains without links corresponds to a column which, without the removed row, is composed of only zeros (for example in the figure \ref{fig:pack_ext_area}a, the third column gets extinct).
Let us call $e_1$ the fraction of extinct columns after one removed row:
\begin{equation}
	e_1 = \frac{1}{M} \sum_{j=1}^{M} \delta_{l_j, 1}
\end{equation}
where the Kronecker delta is equal to $1$ if $l_j = 1$, meaning that all the other elements of the column are zeros.
Iterating this procedure, the fraction of extinct columns at the $k$-th removal step is:
\begin{equation}
	e_k = e_{k-1} + \frac{1}{M} \sum_{j=1}^{M} \delta_{l_j, k} =
	\frac{1}{M}  \sum_{j=1}^{M} \theta(k - l_j)
	\label{eq:ext_curve}
\end{equation}
which is the fraction at the previous step, plus the new extinct columns.
The Heaviside theta function is $1$ only if $l_j \le k$, i.e. the last non-zero element is lesser or equal than the number of removed columns $k$.
The sequence of $e_k$ from $k=1$ to $N$ defines the extinction curve, implying that the extinction area reads:
\begin{equation}
	E = \frac{1}{N} \sum_{k=1}^{N} e_k = \frac{1}{N M} \sum_{j=1}^{M} (N - l_j + 1) = \frac{Z}{N M} + \frac{1}{N}
\end{equation}
where we have used the expression \ref{eq:ext_curve} and the fact that $\sum_{i}^{N} \theta (i - l_j) = N - l_j + 1$, establishing a connection between the extinction area and the area of contiguous zeros.
Note that the term $1/N$ becomes negligible for large matrices, implying the proportionality between $E$ and $Z$.
By transposing the matrix, one can obtain the same relation between the extinction area removing the columns and the area of consecutive zeros from the right side of the matrix.\\

\section{Comparison with the Mariani map}

The map generalization proposed in \cite{mariani2015measuring} reads as follows:
\begin{equation}
	\left\{ \begin{array}{ll}
		\tilde{f}^{(t)}_i = \sum_j A_{ij} q^{(t-1)}_j & \hspace{0.5cm} f^{(t)}_i = \frac{\tilde{f}^{(t)}_i}{\langle \tilde{f}^{(t)} \rangle} 
		\\
		\tilde{q}^{(t)}_j = \left( \sum_i  \frac{A_{ij}}{\left( f^{(t-1)}_i \right)^{\beta} }  \right)^{-\frac{1}{\beta}} &\hspace{0.5cm} q^{(t)}_j = \frac{\tilde{q}^{(t)}_j}{\langle \tilde{q}^{(t)} \rangle} 
	\end{array} \right.
	\label{eq:mar_pn_map}
\end{equation}
which is actually different from our symmetric generalization. This can be seen by imposing the substitution $\tilde{f}_i = \tilde{x}_i$ and $\tilde{q}_j = \tilde{y}_j^{-1/\beta}$.
\begin{equation*}
	\left\{ \begin{array}{ll}
		\tilde{x}^{(t)}_i = \sum_j A_{ij} \left( y^{(t-1)}_j \right)^{-1/\beta} & \hspace{0.5cm} x^{(t)}_i = \frac{\tilde{x}^{(t)}_i}{\langle \tilde{x}^{(t)} \rangle} 
		\\
		\tilde{y}^{(t)}_j = \sum_i  A_{ij} \left( x^{(t-1)}_i\right)^{-\beta}  & \hspace{0.5cm} y^{(t)}_j = \frac{\tilde{y}^{(t)}_j}{\langle \tilde{y}^{(t)} \rangle} 
	\end{array} \right.
\end{equation*}

Indeed they find different node rankings, and therefore different extinction areas, as shown in Figure \ref{fig:mariani} as a function of the map exponent $\gamma$.
In particular, it seems that the symmetric generalization finds always a larger or equal extinction area maximum, and the two maximum coincides only if the symmetric map peak is at $\gamma = 1$, Figure \ref{fig:mariani} \textit{(b)}.

\begin{figure} [h!]
	\centering
	\includegraphics[width=\textwidth]{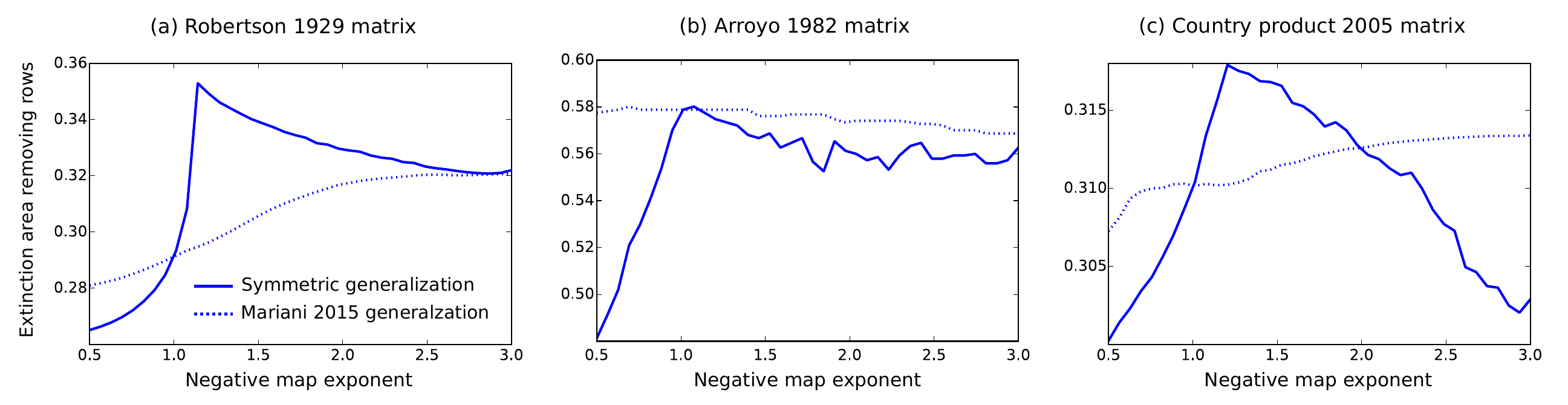}
	\caption{Extinction area computed through the Mariani generalization \cite{mariani2015measuring} and the symmetric one, for three different datasets.
		Consistently for $\gamma = -1$ both the algorithms are equivalent to the standard one and find the same value of extinction area.}
	\label{fig:mariani}
\end{figure}

\clearpage

\begin{table}[!ht]
	\centering
	\resizebox{0.78\textwidth}{!}{
		\begin{tabular}{|l|l|l|l|l|l|l||l|l|l|l|l|l|l|}
			\hline
			name & $n_{rows}$ & $n_{cols}$ & density & $\gamma_{EA}$ & $\gamma_{T}$ & $Z_T$ & name & $n_{rows}$ & $n_{cols}$ & density & $\gamma_{EA}$ & $\gamma_{T}$ & $Z_T$ \\ \hline
			M\_PL\_022 & 21 & 45 & 0.088 & -1.014 & -0.924 & -2.439 & M\_PL\_020 & 20 & 91 & 0.104 & -1.048 & -0.773 & -0.773 \\ \hline
			M\_SD\_022 & 207 & 110 & 0.049 & -1.027 & -0.788 & -5.336 & M\_PL\_028 & 41 & 139 & 0.066 & -1.043 & -0.818 & -0.818 \\ \hline
			A\_HP\_029 & 15 & 34 & 0.155 & -1.109 & -0.894 & -0.698 & M\_PL\_072\_02 & 33 & 63 & 0.082 & -1.006 & -0.833 & -0.833 \\ \hline
			M\_PL\_031 & 48 & 49 & 0.066 & -1.170 & -0.864 & -2.373 & A\_HP\_006 & 16 & 37 & 0.208 & -1.161 & -0.848 & -0.848 \\ \hline
			M\_PL\_030 & 28 & 53 & 0.073 & -1.018 & -0.894 & -0.580 & M\_PL\_017 & 25 & 79 & 0.151 & -1.172 & -0.894 & -0.894 \\ \hline
			M\_PL\_001 & 84 & 101 & 0.043 & -1.073 & -0.803 & -3.675 & A\_HP\_046 & 17 & 39 & 0.305 & -1.240 & -0.667 & -0.667 \\ \hline
			M\_PL\_056 & 91 & 365 & 0.026 & -0.994 & -0.773 & -6.085 & M\_PL\_060\_07 & 29 & 39 & 0.095 & -1.067 & -0.879 & -0.879 \\ \hline
			M\_PL\_051 & 14 & 90 & 0.130 & -1.132 & -0.947 & -1.946 & M\_PL\_060\_17 & 17 & 35 & 0.168 & -1.009 & -0.955 & -0.955 \\ \hline
			M\_PL\_018 & 39 & 105 & 0.094 & -1.044 & -0.879 & -2.582 & FW\_012\_02 & 45 & 33 & 0.100 & -1.033 & -0.818 & -0.818 \\ \hline
			M\_PL\_014 & 29 & 81 & 0.076 & -1.049 & -0.803 & -3.333 & M\_PL\_010 & 31 & 76 & 0.194 & -1.124 & -0.924 & -0.924 \\ \hline
			M\_PL\_058 & 32 & 81 & 0.123 & -1.109 & -0.727 & -2.837 & M\_SD\_004 & 34 & 20 & 0.140 & -1.225 & -1.025 & -1.025 \\ \hline
			M\_SD\_020 & 25 & 33 & 0.182 & -1.150 & -0.811 & -3.553 & FW\_013\_04 & 54 & 46 & 0.108 & -1.157 & -0.818 & -0.818 \\ \hline
			M\_PL\_060\_09 & 18 & 40 & 0.108 & -1.003 & -0.833 & -1.246 & M\_PL\_060\_18 & 20 & 28 & 0.134 & -1.102 & -1.220 & -1.220 \\ \hline
			A\_HP\_031 & 25 & 31 & 0.225 & -1.132 & -0.167 & -0.113 & FW\_010 & 36 & 35 & 0.197 & -1.314 & -0.833 & -0.833 \\ \hline
			FW\_014\_02 & 72 & 41 & 0.191 & -1.124 & -0.818 & -2.457 & FW\_013\_01 & 46 & 36 & 0.113 & -1.159 & -0.848 & -0.848 \\ \hline
			M\_PL\_044 & 110 & 609 & 0.017 & -0.991 & -0.742 & -6.846 & M\_PL\_060\_05 & 33 & 54 & 0.081 & -1.064 & -0.833 & -0.833 \\ \hline
			FW\_013\_03 & 50 & 53 & 0.091 & -1.116 & -0.848 & -3.223 & M\_SD\_034 & 33 & 88 & 0.144 & -1.006 & -0.864 & -0.864 \\ \hline
			M\_PL\_029 & 49 & 118 & 0.060 & -1.021 & -0.818 & -4.901 & M\_SD\_010 & 50 & 14 & 0.334 & -1.264 & -0.530 & -0.530 \\ \hline
			M\_PL\_003 & 36 & 25 & 0.090 & -1.126 & -0.803 & -2.911 & M\_PL\_005 & 96 & 275 & 0.035 & -1.003 & -0.758 & -0.758 \\ \hline
			M\_PL\_060\_15 & 14 & 37 & 0.253 & -1.206 & -0.955 & -0.248 & M\_PA\_004 & 41 & 48 & 0.144 & -1.080 & -0.864 & -0.864 \\ \hline
			M\_SD\_013 & 36 & 19 & 0.288 & -1.109 & -0.758 & 0.423 & M\_PL\_035 & 61 & 36 & 0.081 & -1.018 & -0.833 & -0.833 \\ \hline
			M\_PL\_054 & 113 & 318 & 0.022 & -1.002 & -0.773 & -5.306 & M\_PL\_072\_05 & 17 & 67 & 0.121 & -1.029 & -0.939 & -0.939 \\ \hline
			FW\_014\_01 & 71 & 37 & 0.143 & -1.027 & -0.864 & -2.325 & A\_PH\_005 & 54 & 24 & 0.133 & -1.055 & -0.864 & -0.864 \\ \hline
			M\_SD\_012 & 35 & 29 & 0.144 & -1.061 & -0.909 & -2.037 & M\_PL\_060\_16 & 17 & 39 & 0.172 & -1.160 & -0.848 & -0.848 \\ \hline
			M\_PL\_072\_01 & 39 & 67 & 0.072 & -1.027 & -0.803 & -2.708 & M\_PL\_048 & 30 & 236 & 0.095 & -1.003 & -0.788 & -0.788 \\ \hline
			M\_PL\_052 & 15 & 39 & 0.157 & -1.090 & -0.939 & -0.786 & M\_PL\_021 & 91 & 677 & 0.019 & -0.979 & -0.727 & -0.727 \\ \hline
			M\_PL\_041 & 31 & 43 & 0.109 & -1.058 & -0.788 & -1.541 & M\_PL\_004 & 12 & 102 & 0.136 & -1.067 & -0.985 & -0.985 \\ \hline
			M\_PL\_047 & 19 & 186 & 0.120 & -1.195 & -0.803 & -3.846 & A\_PH\_006 & 6 & 88 & 0.220 & -1.200 & -0.280 & -0.280 \\ \hline
			FW\_014\_03 & 78 & 52 & 0.175 & -1.500 & -0.894 & -4.224 & M\_PL\_023 & 23 & 72 & 0.075 & -1.028 & -0.788 & -0.788 \\ \hline
			M\_PL\_040 & 29 & 43 & 0.091 & -1.170 & -0.879 & -2.918 & FW\_004 & 31 & 28 & 0.161 & -1.285 & -1.121 & -1.121 \\ \hline
			FW\_015\_04 & 94 & 57 & 0.180 & -1.155 & -0.742 & -2.112 & M\_PL\_060\_04 & 21 & 46 & 0.139 & -1.073 & -0.879 & -0.879 \\ \hline
			FW\_003 & 25 & 26 & 0.195 & -1.236 & -0.833 & -2.234 & FW\_011 & 72 & 57 & 0.084 & -1.079 & -0.788 & -0.788 \\ \hline
			A\_PH\_004 & 52 & 22 & 0.161 & -1.252 & -1.136 & -1.661 & A\_HP\_027 & 17 & 30 & 0.212 & -1.079 & -0.114 & -0.114 \\ \hline
			M\_SD\_018 & 29 & 32 & 0.071 & -1.018 & -0.909 & -2.013 & M\_PL\_043 & 28 & 82 & 0.109 & -1.012 & -0.833 & -0.833 \\ \hline
			FW\_013\_02 & 46 & 42 & 0.124 & -0.997 & -0.818 & -3.715 & M\_PL\_009 & 24 & 118 & 0.085 & -1.028 & -0.864 & -0.864 \\ \hline
			M\_PL\_072\_03 & 38 & 79 & 0.080 & -1.055 & -0.818 & -3.297 & M\_PL\_053 & 99 & 294 & 0.020 & -0.983 & -0.773 & -0.773 \\ \hline
			FW\_005 & 39 & 40 & 0.265 & -1.247 & -0.606 & -1.631 & FW\_016\_01 & 37 & 31 & 0.211 & -1.318 & -0.561 & -0.561 \\ \hline
			A\_HP\_010 & 18 & 31 & 0.158 & -1.088 & -0.939 & -1.868 & A\_HP\_050 & 27 & 35 & 0.239 & -1.064 & -0.583 & -0.583 \\ \hline
			A\_HP\_033 & 22 & 25 & 0.360 & -1.125 & -0.523 & -1.149 & M\_PL\_025 & 13 & 44 & 0.250 & -1.012 & -0.818 & -0.818 \\ \hline
			FW\_006 & 30 & 27 & 0.283 & -1.248 & -0.818 & -1.967 & M\_SD\_007 & 72 & 7 & 0.284 & -1.200 & -0.939 & -0.939 \\ \hline
			M\_PL\_062 & 456 & 1044 & 0.032 & -1.197 & -0.485 & -4.410 & M\_PL\_039 & 17 & 51 & 0.149 & -1.194 & -1.439 & -1.439 \\ \hline
			FW\_013\_05 & 82 & 40 & 0.192 & -1.252 & -0.773 & -3.520 & M\_SD\_021 & 18 & 28 & 0.256 & -1.235 & -1.091 & -1.091 \\ \hline
			FW\_014\_04 & 92 & 49 & 0.185 & -1.239 & -0.803 & -4.076 & M\_PL\_006 & 17 & 61 & 0.141 & -1.110 & -0.788 & -0.788 \\ \hline
			FW\_015\_03 & 76 & 43 & 0.165 & -1.048 & -0.803 & -3.234 & M\_PL\_026 & 105 & 54 & 0.036 & -1.095 & -0.788 & -0.788 \\ \hline
			M\_PL\_012 & 29 & 55 & 0.091 & -1.253 & -0.947 & -2.943 & A\_HP\_042 & 21 & 32 & 0.125 & -0.997 & -0.818 & -0.818 \\ \hline
			M\_SD\_016 & 24 & 61 & 0.342 & -1.184 & -0.652 & -0.886 & M\_PL\_002 & 43 & 64 & 0.071 & -1.012 & -0.833 & -0.833 \\ \hline
			M\_PL\_060\_03 & 13 & 45 & 0.222 & -1.042 & -0.788 & -1.708 & M\_PL\_013 & 9 & 56 & 0.204 & -1.081 & -1.333 & -1.333 \\ \hline
			M\_PL\_055 & 64 & 195 & 0.035 & -1.012 & -0.803 & -3.687 & M\_PL\_034 & 26 & 128 & 0.094 & -1.039 & -0.848 & -0.848 \\ \hline
			A\_HP\_044 & 27 & 26 & 0.281 & -1.101 & -1.061 & -0.821 & FW\_007 & 39 & 42 & 0.135 & -1.073 & -0.864 & -0.864 \\ \hline
			M\_PL\_046 & 16 & 44 & 0.395 & -1.082 & -0.621 & -1.971 & FW\_012\_01 & 38 & 30 & 0.111 & -1.055 & -0.788 & -0.788 \\ \hline
			M\_PL\_049 & 37 & 225 & 0.071 & -1.033 & -0.833 & -4.899 & M\_PL\_060\_06 & 26 & 45 & 0.082 & -1.018 & -0.864 & -0.864 \\ \hline
			M\_PL\_027 & 18 & 60 & 0.111 & -1.009 & -0.894 & -1.055 & A\_HP\_025 & 18 & 40 & 0.149 & -1.098 & -0.864 & -0.864 \\ \hline
			M\_PL\_007 & 16 & 36 & 0.148 & -1.127 & -0.970 & -2.028 & FW\_009 & 37 & 37 & 0.176 & -1.412 & -1.242 & -1.242 \\ \hline
			FW\_008 & 248 & 244 & 0.055 & -1.088 & -0.742 & -12.006 & FW\_015\_01 & 76 & 49 & 0.170 & -1.197 & -0.788 & -0.788 \\ \hline
			M\_PL\_019 & 40 & 85 & 0.078 & -0.997 & -0.818 & -3.017 & M\_PL\_060\_08 & 19 & 28 & 0.135 & -1.201 & -1.015 & -1.015 \\ \hline
			M\_PL\_016 & 26 & 179 & 0.089 & -1.121 & -0.864 & -3.225 & M\_SD\_019 & 169 & 40 & 0.099 & -1.064 & -0.864 & -0.864 \\ \hline
			M\_PL\_072\_04 & 25 & 69 & 0.111 & -1.107 & -1.273 & -2.525 & FW\_015\_02 & 66 & 37 & 0.170 & -1.052 & -1.318 & -1.318 \\ \hline
			M\_PL\_057 & 114 & 883 & 0.019 & -0.997 & -0.727 & -7.025 & M\_PL\_015 & 131 & 666 & 0.034 & -0.997 & -0.803 & -0.803 \\ \hline
	\end{tabular}}
	\caption{List of the ecological matrices considered for the analysis of the extinction area and nestedness. The names refer to the identifier in the Web of Life database. We count only the rows and columns with at leas one element and the density is computed on the binarized matrix.
		The exponents of the map that maximize the extinction area, the one that minimize the nested temperature and the Z-score of nested temperature are also reported.}
\end{table}

\begin{figure}
	\centering
	\includegraphics[width=0.8\linewidth]{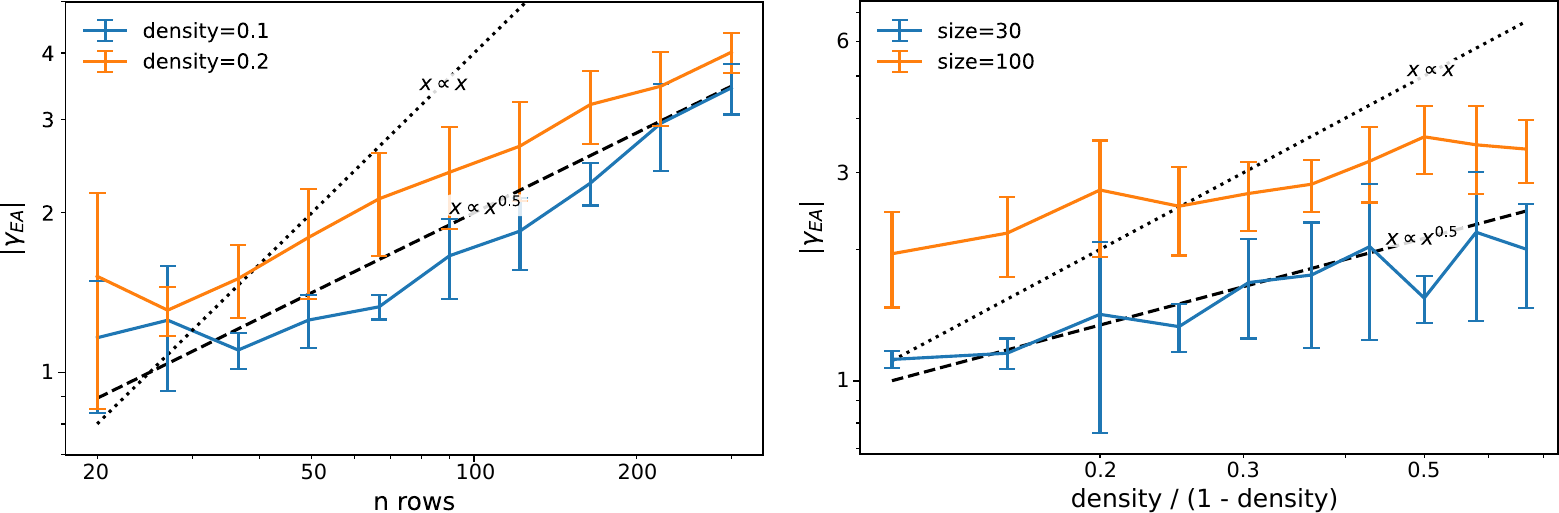}
	\caption{
		{Exponents maximizing the extinction-area for different matrix properties of uniform random matrices.
			Each point is the average over an ensemble of 10 matrices at fixed property (i.e. size or density) whose standard deviation defines the error bar. 
			On the left, the absolute value of the exponent is plotted as a function of the number of rows of the squared random matrix for two values of densities.
			On the right, the x-axis  represents the density as $d / (1-d)$, the same variable used in Figure 4b of the main text.
			Despite the big variability shown by the error bars, the scaling of the average is compatible with the square root of the two properties on the x-axes, which is the one predicted for the transition exponent of the map.}
	}
	\label{fig:EA_rc}
\end{figure}

\begin{figure}
	\centering
	\includegraphics[width=0.8\linewidth]{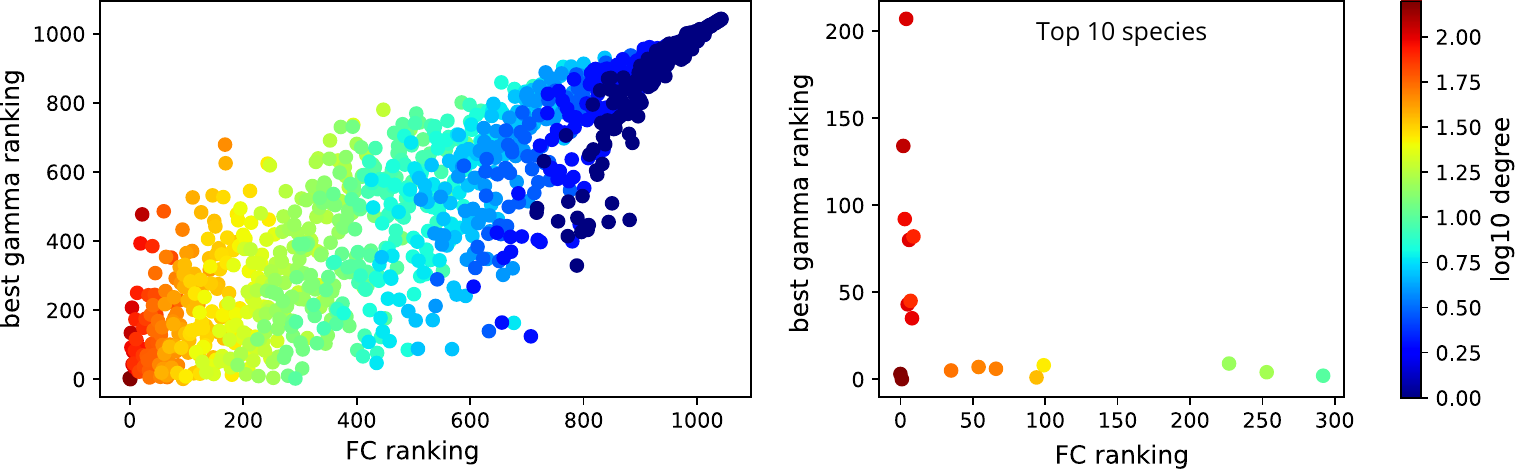}
	\caption{
		{Scatter plot of the ranking given by the fitness-complexity map and the ranking given by our map with the $\gamma$  maximizing  the extinction area in the mutualistic system Robertson 1929. Each point is a species, whose ranking according to the two algorithms is on the axis.
			The colors are the $\log_{10}$ of the species degree.
			On the right,  we select species which are ranked in the top-10 in at least one of the two rankings.}
	}
	\label{fig:EA_rc}
\end{figure}

\begin{figure}
	\centering
	\includegraphics[width=0.6\linewidth]{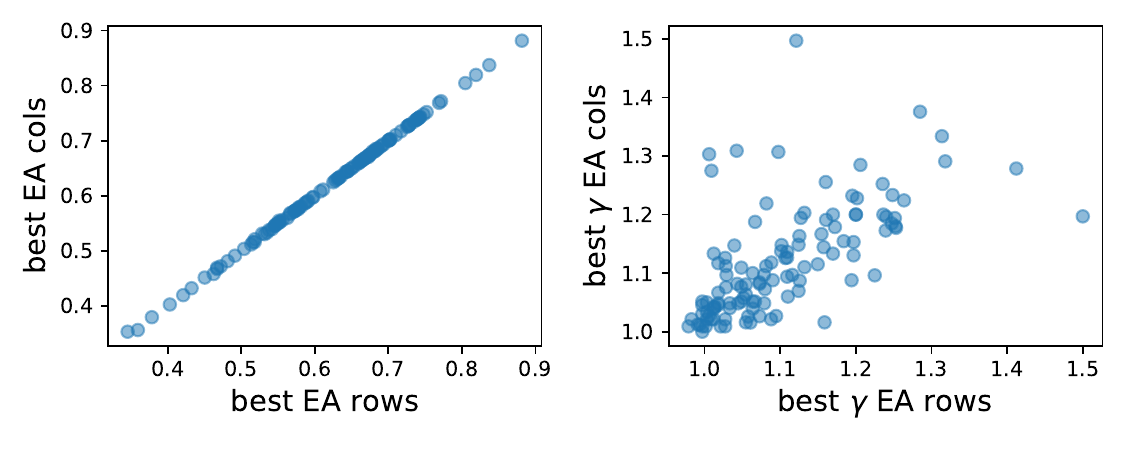}
	\caption{{\textit{(Left)} Perfect correlation between the maximal  extinction area values obtained by  removing rows or removing columns found by our algorithm.
			\textit{(Right)}: Correlation between the  exponents that our map finds in optimizing for the extinction area by removing rows or columns. Spearman correlation $0.71$, p-val $4\;10^{-19}$.}
	}
	\label{fig:EA_rc}
\end{figure}

\begin{figure}
	\centering
	\includegraphics[width=\linewidth]{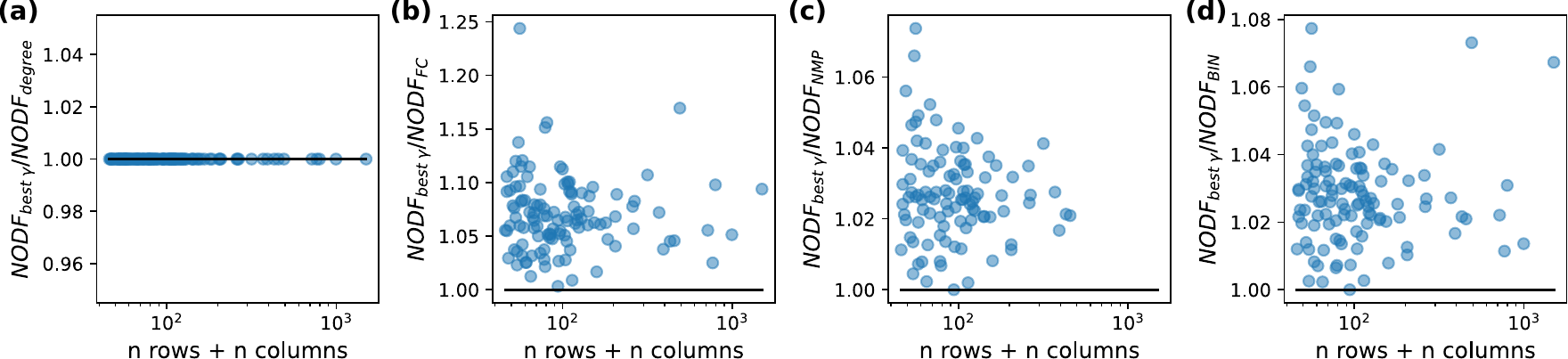}
	\caption{{Comparison of the NODF metric for the nestedness computation as a ratio between the score of our algorithm (at the value of $\gamma$ that maximizes the score) and four other ways of ordering the matrix.
			From the left to the right, the other rankings are provided by the degree of rows and columns (a), the Fitness Complexity map (b), the algorithm proposed in \cite{mariani2024ranking} (c) and the BINMATNEST algorithm (d).}}
	\label{fig:NODF}
\end{figure}

\begin{figure}
	\centering
	\includegraphics[width=1\linewidth]{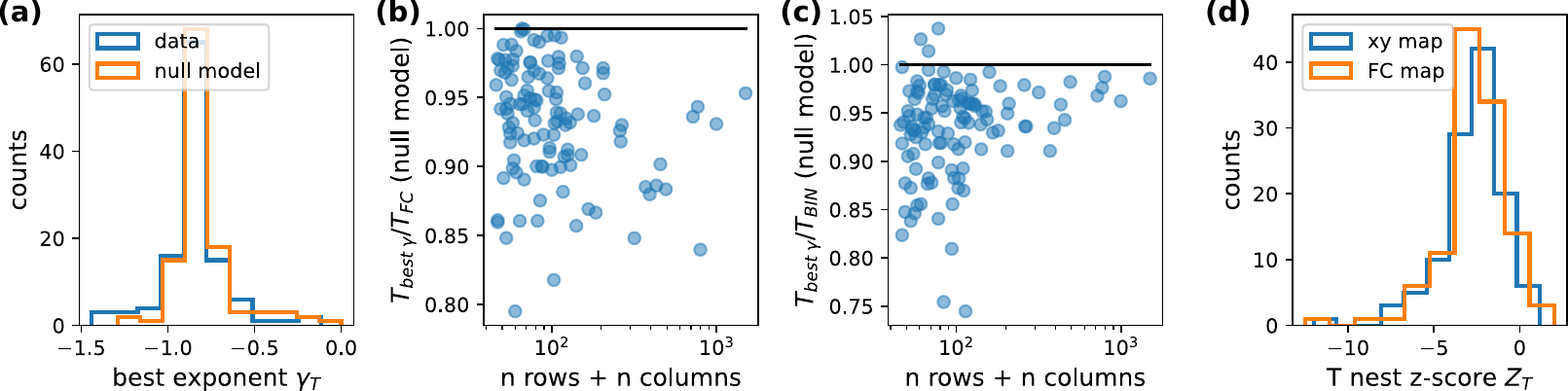}
	\caption{{Panel (a): Histogram of the exponents at which the map maximizes the nestedness in the real ecological matrices and in random counterparts that preserve the degree on average.
			Panel (b,c) report  for the degree-preserving random matrices the  ratio of the nestedness temperature for the best-gamma ranking and a second algorithm: fitness complexity (b) and BINMATNEST (c).
			Panel (d): histogram of Z-score for the nestedness temperature computed by our map or by the fitness-complexity map.}}
	\label{fig:NODF}
\end{figure}

\end{document}